\documentclass[a4paper,10pt]{article}

\usepackage{latexsym}
\usepackage{amsmath,amsfonts,amssymb,amsthm}
\usepackage{graphicx,color,epsfig}

\def\R{\mathbb{R}}
\def\N{\mathbb{N}}
\def\Z{\mathbb{Z}}

\begin{document}

\noindent
{\LARGE \bf Selective chaos of travelling waves in feedforward chains of bistable maps}
\bigskip

\noindent
{\large \bf Bastien Fernandez}
\bigskip

\noindent
Laboratoire de Probabilit\'es, Statistique et Mod\'elisation, CNRS - Univ.\ Paris Denis Diderot - Sorbonne Univ., 75205 Paris CEDEX 13 France
\bigskip

\noindent
{\bf Abstract. We study the chaos of travelling waves (TW) in unidirectional chains of bistable maps. Previous numerical results suggested that this property is selective, {\sl viz.}\ given the parameters, there is at most a single (non-trivial) velocity for which the corresponding set of wave profiles has positive topological entropy. However, mathematical proofs have remained elusive, in particular because the related symbolic dynamics involves entire past sequences. Here, we consider instead inite (short) rank approximations for which the symbolic dynamics has finite memory. For every possible velocity, we compute the existence domains of all possible finite type subshifts of TW with positive entropy. In all examples, chaos of TW turns out to be selective, indeed.}
\bigskip

\noindent
{\em Dedicated to the memory of Valentin Afraimovich.}

\section{Introduction}
Valentin Afraimovich had a strong interest for lattices of coupled dynamical systems, the so-called Lattice Dynamical Systems (LDS). He made a number of diverse and important theoretical contributions to this field; a summary of them can be found in the notes of his lecture at the CML2004 school in Paris \cite{A05}. In short terms, a LDS is a (continuous or discrete time) dynamical system whose phase space is ${\cal M}^{\Z^d}$ (lattice configurations), where ${\cal M}$ is a subset of $\R$ or of a (compact) manifold, and $d\in\N$ is the lattice dimension.

In this setting, Valentin introduced me to the problem of the detection of a {\em preferred direction in space-time} (from his own words, often accompanied with expressive hand motions!), presumably a measure of the velocity of information flow in the system. During several years, we frequently spent time together in various places, San Luis Potosi, Marseille, etc. While he continuously showed receptive patience, Valentin was moved by a strong will, which manifested itself as a vigorous stimulation. A young fellow at that time, I had no previous experience of interaction with a senior colleague who was both scientifically demanding and open to feedback and discussion. This collaborative experience has been truly beneficial to me, and also a particularly good time of a close relationship. 

Valentin's incentive brought us (also with Antonio Morante) to extend Milnor's notion of directional entropy in cellular automata \cite{M86,M88} to lattice dynamical systems \cite{ACFM02}. Two context-dependent definitions emerged. The first one involved space-time normalisation and was intended for systems with chaotic (temporal) dynamics. The second one used temporal normalisation only and was aimed at the case of regular dynamics. 

We proceeded to an extended investigation of basic characteristics, by analogy with the analysis of the topological entropy in dynamical systems \cite{R99}. Furthermore, we obtained explicit estimates in simple examples, which unexpectedly showed that the space-time normalised quantity did not depend on the direction. This invariance was later confirmed to hold in every translation invariant LDS,\footnote{Translation invariant means that the dynamics commutes with the operator $\sigma$ of spatial translations, which is defined below.} thanks to a decisive contribution by Edgardo Ugalde \cite{AMU03}.\footnote{More precisely, the entropy as defined in \cite{ACFM02} depended on the direction. However, \cite{AMU03} showed that this dependence was trivial, because it could be removed by suitable renormalisation.} Therefore, the directional entropy unfortunately appears to be insufficient to detect preferred space-time directions, at least for chaotic systems whose dynamics commutes with spatial translations. 

What about systems with regular dynamics? The space-time normalised directional entropy would have to vanish but not necessarily the time normalised one. Moreover, the examples below \cite{ACFM02} show non trivial dependence and call for further investigation.   

\section{Unidirectional systems, chaos of travelling waves and symbolic dynamics}
The probably simplest (non trivial) example of LDS with regular dynamics is the direct product (over $\Z$) of bistable one-dimensional maps ({\sl ie.}\ maps of the interval having two stable fixed points). Explicitly, the dynamics of configurations $x\in [0,1]^\Z$ is generated by the map $F$ defined by
\[
(F (x))_s=f(x_s),\ \forall s\in\Z.
\]
Assuming that $f(0)=0$ and $f(1)=1$ are the two stable fixed points of $f$, any configuration in $\{0,1\}^\Z$ must be a (stable) fixed point of $F$. This property is an instance of {\sl spatial chaos} in LDS,\footnote{Another area of Valentin's expertise, a LDS is said to have spatial chaos when the action of the spatial translations $(\sigma (x))_s=x_{s-1}$ on the set of LDS fixed points (viewed as a dynamical system whose time is given by the spatial variable $s$ \cite{CER87,EP91}), is chaotic, typically with positive topological entropy \cite{A05}.} and it implies that the (time normalised) directional entropy vanishes in the direction of time and is maximal in the direction of space.  

The next example is when this uncoupled system is combined with the spatial transition operator $\sigma$, {\sl viz.}\ 
\[
(\sigma \circ F (x))_s=f(x_{s-1}),\ \forall s\in\Z.  
\]
Under the same circumstances, the system $\sigma\circ F$ has a spatial chaos, now of (convectively stable) travelling waves of velocity 1. Its directional entropy vanishes in the direction of wave propagation (corresponding to the angle $\tfrac{\pi}{4}$ in the space-time) and, as before, is maximal in the orthogonal direction. 

These examples are somewhat naive and rather irrelevant. As modelling of transport phenomena is concerned, it would be more interesting to have similar  information for the following unidirectional chain of coupled (bistable) maps
\begin{equation}
(F_\epsilon (x))_s=(1-\epsilon) f(x_s)+\epsilon  f(x_{s-1}),\ \forall s\in\Z, 
\label{UNIDIRCHAIN}
\end{equation}
for $\epsilon\in [0,1]$ (NB: this system reduces to the previous examples for the limit values $\epsilon=0$ and $\epsilon=1$ respectively).
So far, estimates of the directional entropy for this chain have remained elusive (expected in the neighbourhood of the limit cases, using perturbative arguments). To address this issue, the considerations in the limit cases above suggest to investigate the chaos of travelling waves (TW) for an arbitrary velocity. 

Travelling waves in (continuous or discrete time) LDS can be defined following the basic notion in physics, namely that they are trajectories given by $x_s^t=u(s-vt)$ for some profile $u:\R\to\R$ and some velocity $v\in\R$ \cite{CF97a}. However, TW in discrete-time LDS can also be characterised using spatial translations \cite{AP93}, including for irrational velocities \cite{AF00}. In particular, any solution of the equation  
\begin{equation}
G^q(x)=\sigma^p(x),
\label{TWRAT}
\end{equation}
where $p\in\Z$ and $q\in\N$, defines a TW of velocity $v=\tfrac{p}{q}$ for the LDS generated by the map $G$. Of note, the velocity is constrained by the coupling range, in particular we must have $v\in [0,1]$ for the LDS \eqref{UNIDIRCHAIN}.

Extending the notion above, the LDS generated by $G$ is said to have (spatial) {\sl chaos of TW} of velocity $\tfrac{p}{q}$ if the action of $\sigma$ on the set of solutions of \eqref{TWRAT} has positive topological entropy. Together with Vladimir Nekorkin, Valentin has asserted the existence of chaos of TW in some examples of LDS by constructing horseshoes of the profile generating dynamics associated with equation \eqref{TWRAT} \cite{AN94}. This approach can be viewed as a nonlinear extension of the transfer matrix technique in theoretical solid-state physics. However, the construction is velocity specific and makes it difficult to evaluate the velocity dependence on parameters, not to mention to portray a global description of the chaos of TW in parameter space. 

\begin{figure}[ht]
\begin{center}
\includegraphics*[width=50mm]{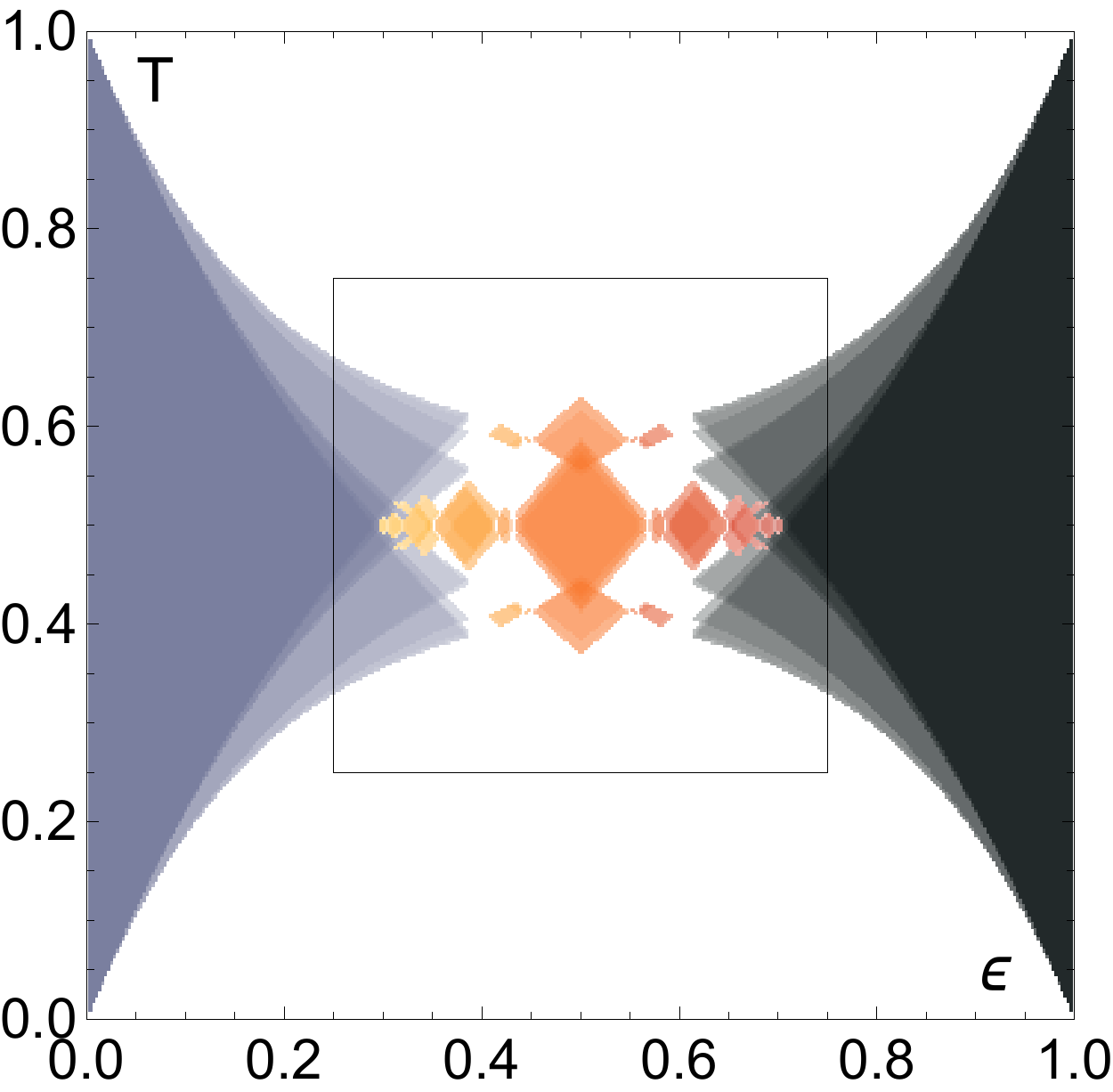}
\hspace{1.5cm}
\includegraphics*[width=50mm]{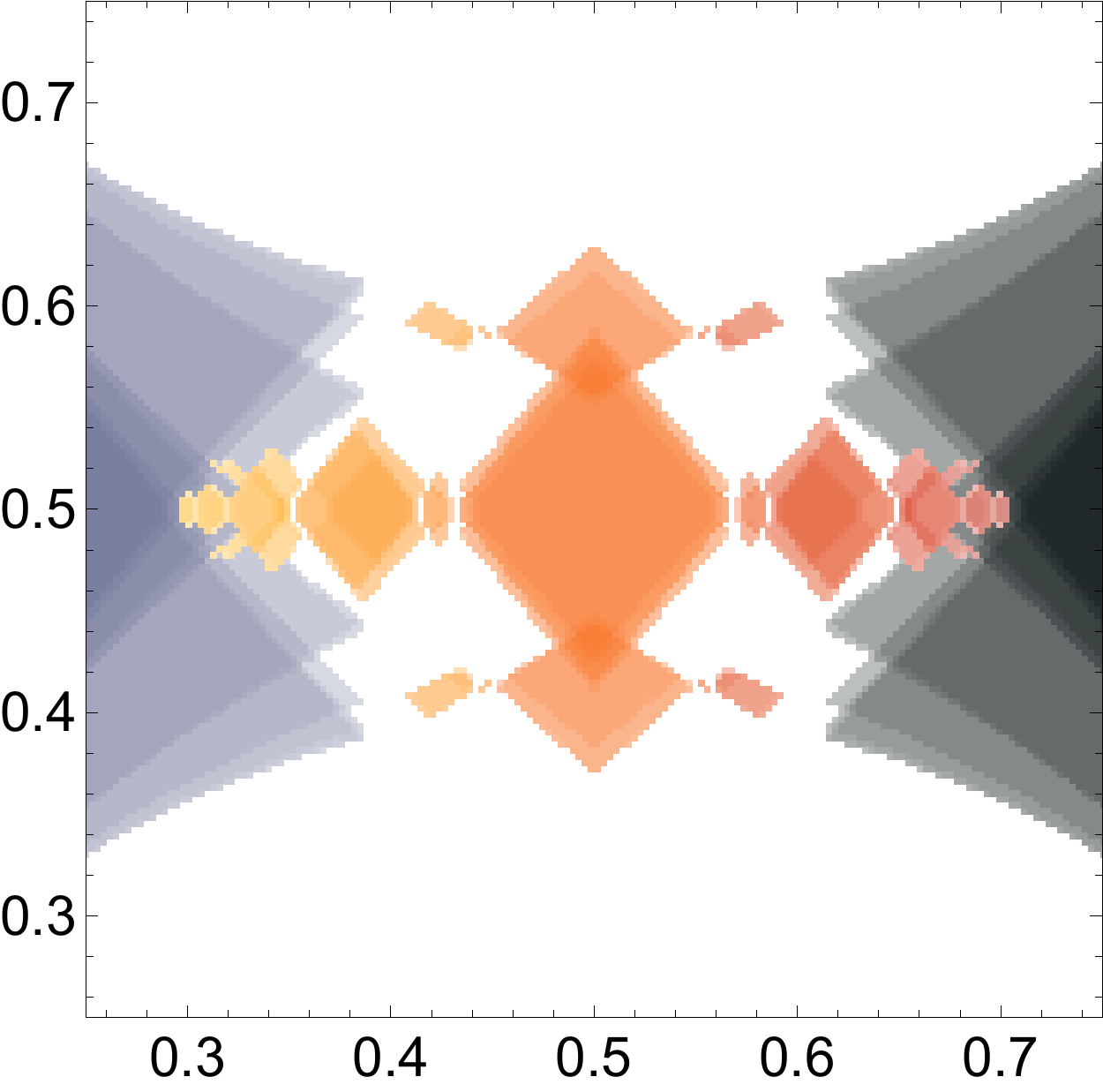}
\end{center}
\caption{Entropy-velocity diagrams of TW of the LDS \eqref{UNIDIRCHAIN} with individual map \eqref{BISTABLEF} for $a = 0.6$ ({\sl Left:} Original diagram in the full square of the parameters $(\epsilon,T)$. {\sl Right:} Zoom into the central region delimited by the square). Painted points represent the velocity dependent quantity $\tfrac{\log P_{13,v}}{13}$ (approximation of the TW entropy), based on color and intensity. The color indicates the velocity $v\in V\cup 1-V$ where $V=\{0,\tfrac16,\tfrac15,\tfrac14,\tfrac13,\tfrac25,\tfrac37,\tfrac12\}$, 0 (resp.\ 1) is painted in gray (resp.\ black) and the other velocity colors monotonically range from yellow to red. The intensity is proportional to the entropy, from 0 to $\log 2$. Positive entropy domain of intermediate velocities $v\in (0,1)$ appear to be pair-wise disjoints. However, the domains with small velocities overlap with the one at $v=0$ (and similarly in the right part of the pictures) indicating that a chaos of fixed points (resp.\ of TW with $v = 1$) may coexist with a chaos of patterns with intermediate velocity.}
\label{ENTROPYVELOCITYDIAG}
\end{figure}
To address this issue, we considered a special case of the unidirectional chain \eqref{UNIDIRCHAIN} above, when the individual map $f$ is piecewise affine with two branches of unique slope, separated by a discontinuity. Formally speaking, the map writes
\begin{equation}
f(u)=au+(1-a)H(u-T),\ \forall u\in [0,1],
\label{BISTABLEF}
\end{equation}
where $H$ is the (right continuous) Heaviside function, the slope $a\in [0,1)$ and the discontinuity $T\in (0,1)$. That this map is a piecewise contraction implies that any TW away from the discontinuity must be convectively stable \cite{CF97a}, and in particular that any chaos of TW must be a chaos of stable TW.

By using symbolic dynamics (see below), we numerically discovered \cite{FLU09,FLU11} that, given any values of the parameters, there is at most one (non-trivial) velocity $v\in (0,1)$ for which the corresponding TW set may have positive entropy,\footnote{In symbolic systems, the topological entropy can be defined as the exponential growth rate of the number of admissible blocks \cite{LM95}. For numerical purposes, following \cite{BP97}, we used instead the number of periodic points and assumed that the entropy of the set of TW profiles with velocity $v$ is given by $\limsup_{L\to +\infty}\tfrac{\log P_{L,v}}{L}$ where $P_{L,v}$ is the number of $L$-periodic blocks of velocity $v$.} see Fig.\ \ref{ENTROPYVELOCITYDIAG} for an illustration for $a=0.6$. Depending on the parameters, either there is no chaos of TW, or we have a chaos of TW with trivial velocity 0 or 1, or such chaos is combined with a chaos of TW with $v\in (0,1)$, but we have never observed two chaos of TW with distinct non-trivial velocities. This observation might not be what Valentin had expected but it can be considered as a selection of a unique, if not preferred, direction in space-time (when one discards the background velocities 0 and 1). 

Besides its obvious symmetric features,\footnote{The symmetric features of Fig.\ \ref{ENTROPYVELOCITYDIAG} are consequence of the following symmetries of the LDS:
\begin{itemize}
\item If $\{x_s^t\}$ is a trajectory for $T$ (with all $x_s^t\neq T$), then $\{1-x_s^t\}$ is a trajectory for $1-T$.
\item If $F_\epsilon^q(x)=\sigma^p(x)$ then $F_{1-\epsilon}^q\circ R(x)=\sigma^{q-p}\circ R(x)$ where $(R(x))_s=x_{-s}$ for all $s$.
\end{itemize}} the numerical results called for rigorous explanation. That a chaos of TW with $v\in (0,1)$ could coexist with one at $v\in\{0,1\}$, combined with proofs of parameter dependent coexistence of TW with different velocities \cite{FLU09}, suggested that the proof could not be elementary. In fact, excepted for the extreme velocities $v=0$ and $v=1$ for which the exact parameter domain for existence of full chaos could be determined \cite{CF97b}, for arbitrary velocities, only estimates of pairwise disjoint domains of existence of chaotic sets of TW were available, based on shadowing arguments. In order to provide further insights into analytic arguments, we need to introduce considerations about symbolic dynamics.

A convenient aspect of the piecewise affine system \eqref{UNIDIRCHAIN} with individual map \eqref{BISTABLEF} is that its symbolic description is easily accessible. Indeed, any trajectory $\{x^t\}$ (where $x^{t+1}=F_\epsilon(x^t)$) can be coded by a space-time symbolic sequence $\{\theta_s^t\}$ via the relation $\theta_s^t=H(x_s^t-T)\in \{0,1\}$. Conversely, and more importantly, symbolic sequences uniquely determine trajectories in the attractor of the LDS. Indeed, solving the iterations associated with trajectories whose components exist and are bounded for all $t\in\Z$ yields the following expression \cite{CF97a,FLU09}
\begin{equation}
x_s^t=(1-a)\sum_{k=1}^\infty a^{k-1}\sum_{n=0}^k\ell_{n,k}\theta_{s-n}^{t-k}
\label{GLOBORB}
\end{equation}
where the coefficients $\ell_{n,k}=\binom{k}{n}(1-\epsilon)^{k-n}\epsilon^n\geq 0$ satisfy the evident normalisation
\[
\sum_{n=0}^k\ell_{n,k}=1,\ \forall k\in\N 
\]
Not only the LDS attractor can be fully specified using symbolic dynamics, but so do the topological properties of the dynamics in this set, when symbolic sequences are endowed with a suitable topology. In particular chaos of TW can be analyzed and quantified in the symbolic context. 

This topological equivalence between the dynamics in the original space and its symbolic representation is standard in the theory of dynamical systems. However, what is specific to the current setting is the following iterative process for symbolic codes 
\begin{equation}
\theta_s^{t+1}=H\left((1-a)\sum_{k=0}^\infty a^{k}\sum_{n=0}^{k+1}\ell_{n,k+1}\theta_{s-n}^{t-k}-T\right),\quad \forall s,t\in\Z
\label{ADMIEQU}
\end{equation}
obtained by stipulating that the code of any trajectory given by \eqref{GLOBORB} must coincide with the input symbolic sequence. In other words, the LDS attractor and its topological properties can be captured by an explicit iteration scheme for symbolic codes. This formulation has proved useful in several cases \cite{CF97a,CF97b,FLU09}, for instance to obtain the estimates mentioned above about chaos of TW.   
 
\section{Selective chaos of travelling waves for finite rank approximations}
A full mathematical proof of the numerical results of \cite{FLU09} remains elusive, especially because the iterations \eqref{ADMIEQU} involve the entire past sequence ({\sl ie.}\ the series in $k$ is infinite when $a>0$) and this makes it virtually impossible to determine all solutions for arbitrary values of the parameters. Here, we suggest to investigate instead the following finite rank iteration schemes, obtained by truncating the series in the expression \eqref{GLOBORB} (and using a suitable normalisation)
\[
\theta^{t+1}={\cal F}_R(\theta^{t-R+1},\cdots,\theta^t)\quad\text{where}\quad \left({\cal F}_R(\theta^{-R+1},\cdots,\theta^0)\right)_s=H\left(\frac{1-a}{1-a^R}\sum_{k=0}^{R-1} a^k\sum_{n=0}^{k+1}\ell_{n,k+1}\theta_{s-n}^{-k}-T\right).
\]
The iteration scheme ${\cal F}_R$ can be viewed as a kind of cellular automaton (CA), which, thanks to the updated normalisation, actually shows the same symmetries as the original LDS. Our goal is to systematically determine the existence domains of TW for increasing values of $R$. Exponential decay of the coefficients $a^{k}$ suggests that the ${\cal F}_R$ should be good approximations of the LDS; the larger $R$, the better the approximation. The dependence on $a$ indicates that the approximation should be better for smaller values of this parameter. 

Since ${\cal F}_R$ only involves input symbols in a time window of length $R$, the velocity of the TW ${\cal F}_R^q(\theta)=\sigma^p(\theta)$ must be of the form $\frac{p}{q}$ for $q\in\{1,\cdots,R\}$ and $p\in\{0,\cdots,q\}$ (and assuming wlog that $p$ and $q$ are co-prime). Similarly, the inputs are limited to the $R+1$ neighbour sites, so the TW profile $\{\theta_s\}$, when read from left to right, can be regarded as being generated by a topological Markov chains (subshift of finite type); given an arbitrary site $s$ and an admissible $(R+1)$-block $\theta_{s-R+1}\cdots\theta_s$, the subshift specifies those subsequents admissible blocks $\theta_{s-R+2}\cdots\theta_{s+1}$ at the next site. This viewpoint is convenient for the computation of the TW entropy, as the logarithm of the largest eigenvalue of the corresponding transition matrix (see \cite{LM95} for more details on finite-type subshift and the computation of their entropy). 
 
\subsection{Rank 1 approximation}
\begin{figure}[ht]
\begin{center}
\includegraphics*[width=50mm]{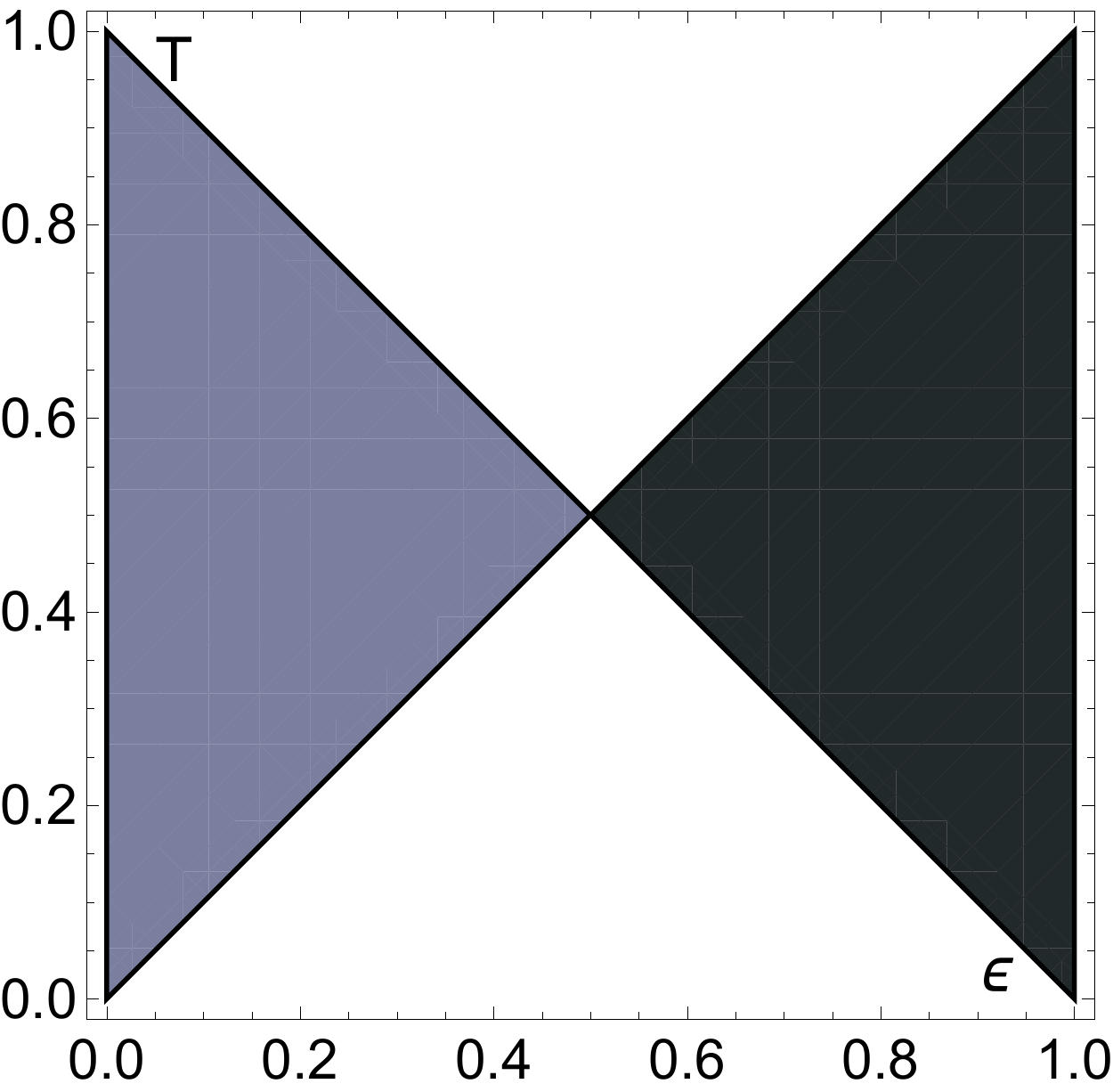}
\end{center}
\caption{Entropy-velocity diagrams of TW of the cellular automaton ${\cal F}_1$. The square of parameters $(\epsilon,T)$ decomposes into 4 regions. In the left (blue) triangle, every element $\theta\in\{0,1\}^\Z$ is invariant under the dynamics (full chaos of fixed points). In the lower white triangle, the only non-homogeneous TW profiles are $\{H(s)\}\ \text{mod}\ \sigma$ ($v=0$) and $\{H(-s)\}\ \text{mod}\ \sigma$ ($v=1$). The TW in the other domains follow from symmetries.}
\label{DIAGAPPROX}
\end{figure}
The simplest approximation is when the summation on $k$ only has one term ($R=1$) so that the iterative process on symbolic sequences reduces to one for elements of $\{0,1\}^\Z$, {\sl viz.}\ we have the following (genuine) CA 
\[
\left({\cal F}_1(\theta)\right)_s=H\left((1-\epsilon) \theta_s+\epsilon \theta_{s-1}-T\right)
\]
The TW velocities of ${\cal F}_1$ are $v\in\{0,1\}$ are trivial. By the symmetry $v \leftrightarrow 1-v$, it suffices to study the fixed points ($v=0$). To that goal, let 
\[
X_1(\theta_{1}\theta_0)=(1-\epsilon) \theta_0+\epsilon \theta_1.
\]
It is simple to check that
\begin{itemize}
\item In the interval $0<T\leq \min\left\{X_1(10),X_1(01)\right\}=\min\left\{\epsilon,1-\epsilon\right\}$, the only possible fixed points are the homogeneous ones $0^\Z$ and $1^\Z$, and $\{H(s)\}\ \text{mod}\ \sigma$.\footnote{All the LDS here are translation invariant; hence all their trajectories come as equivalence classes $\{x^t\}\ \text{mod}\ \sigma$ or $\{\theta^t\}\ \text{mod}\ \sigma$.}  Indeed, we then have $H(X_1(01)-T)=H(X_1(10)-T)=1$. Furthermore, using that $X_1(01)=1-X_1(10)$, a symmetric conclusion holds for $T\in (X_1(01),1]$.
\item In the interval $\epsilon=X_1(10)<T\leq X_1(01)=1-\epsilon$, (which is non-empty obviously iff $\epsilon<\tfrac12$) every element of $\{0,1\}^\Z$ is a fixed point of ${\cal F}_1$ (full chaos of fixed points, with entropy equal to $\log 2$).
\end{itemize}
The entropy-velocity diagram of TW for ${\cal F}_1$ is given in Fig.\ \ref{DIAGAPPROX}. The situation is pretty obvious in this case as the two domains of existence of chaos of TW do not overlap.

\subsection{Rank 2 approximation}
\begin{figure}[ht]
\begin{center}
\includegraphics*[width=50mm]{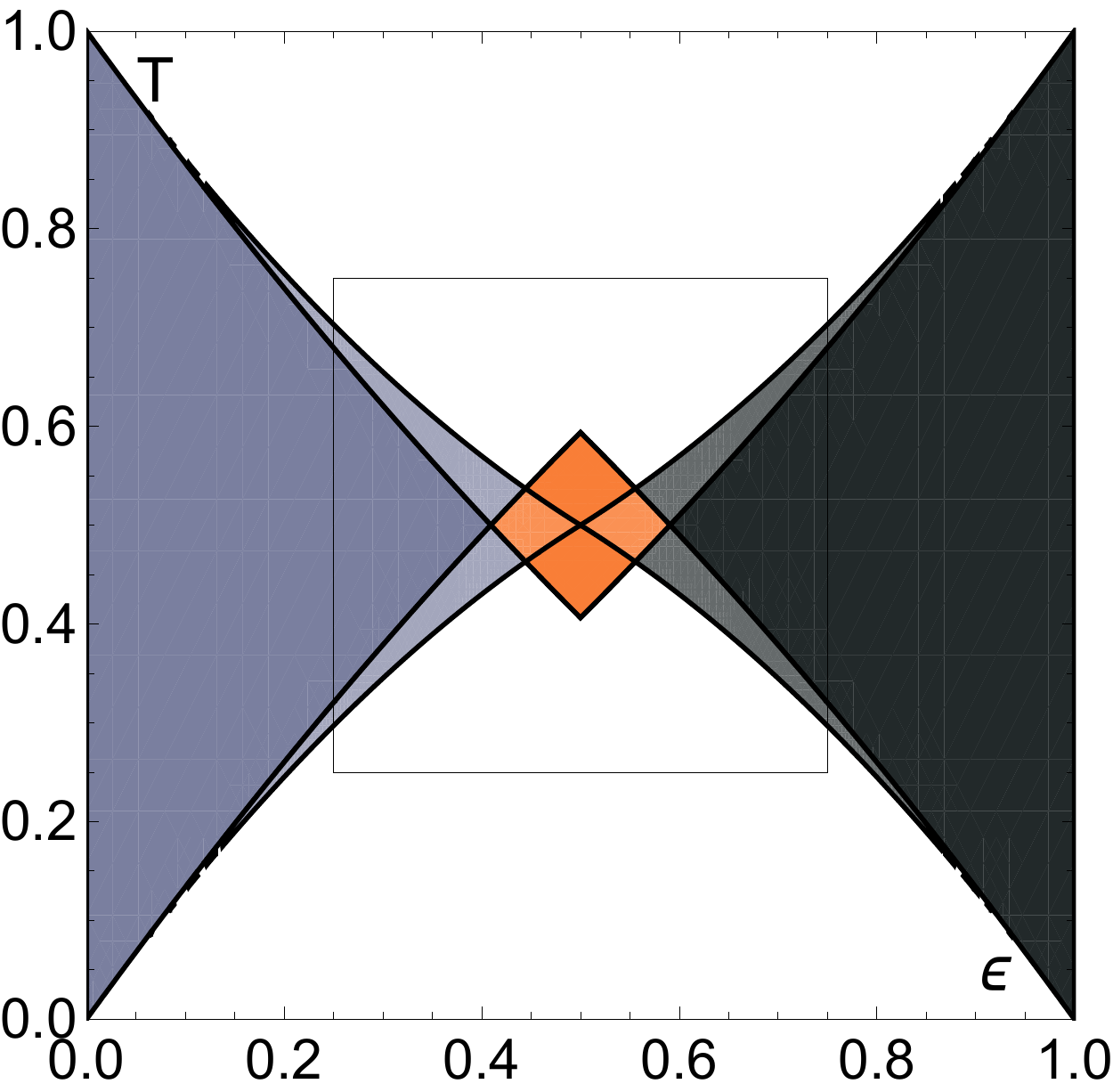}
\hspace{1.5cm}
\includegraphics*[width=50mm]{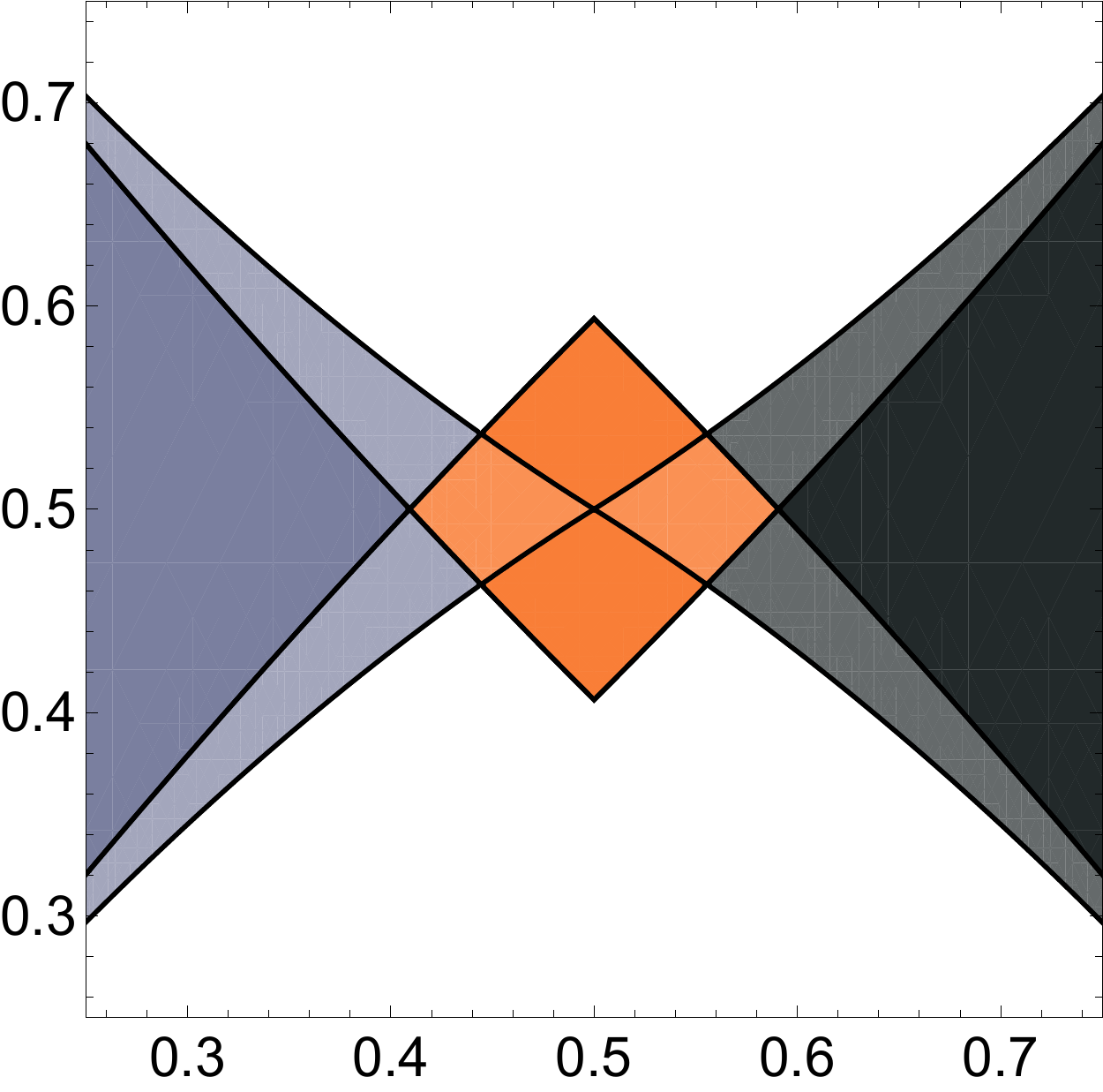}
\end{center}
\caption{Entropy-velocity diagram of TW of the cellular automaton ${\cal F}_2$ (and zoom into the central region). Chaos of fixed points is full in the central blue region (all 3-blocks are admissible and entropy = $\log 2$) and partial in the two other blue domains (either $110$ or $001$ is forbidden, depending on the domain/entropy = $\log\frac{1+\sqrt{5}}2$ in both cases). In the blacks domains, the same comments apply to $v=1$ TW. Orange domains correspond to chaos of TW with $v=\frac12$. In particular, in low intensity domains (left and right), both $010$ and $101$ are forbidden (entropy = $\log\frac{1+\sqrt{5}}2$). In each of the high intensity domains (bottom and top), only one of these words is forbidden (entropy = $\log 1.755$). Finally, the only admissible non-homogeneous profiles in the white domains are the same as in Fig.\ \ref{DIAGAPPROX}.}
\label{APPROXRANK2}
\end{figure}
We now turn to the analysis of waves for ${\cal F}_2$. 
The only possible TW velocities in this case are $v\in\{0,\tfrac12,1\}$. Let 
\[
X_2\left( \begin{matrix}\theta_{2}^{1}\theta_{1}^{1}\theta_0^{1}\\ \hfill\theta_{1}^{0}\theta_0^{0}\end{matrix}\right)=\frac{1}{1+a}\sum_{k=0}^1 a^k\sum_{n=0}^{k+1}\ell_{n,k+1}\theta_{n}^{k},
\]
(where sub- and super-scripts are denoted with non-negative integers for simplicity) which obviously depends on $a$ and $\epsilon$.

\subsubsection{Analysis of fixed points}
\begin{figure}[ht]
\begin{center}
\includegraphics*[height=35mm]{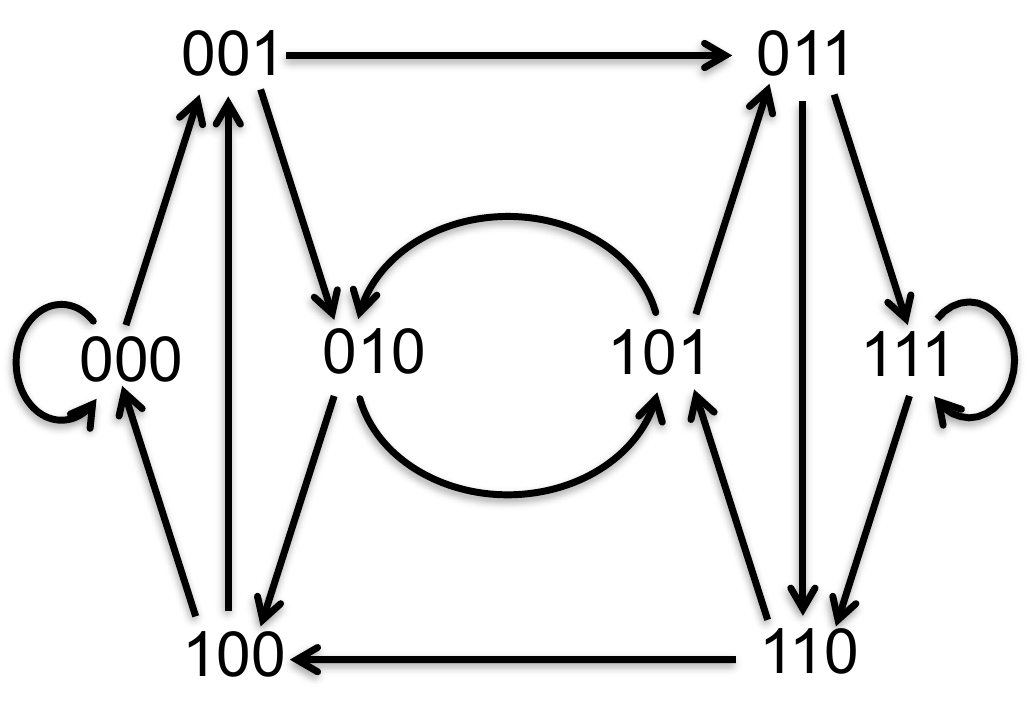}
\hspace{0.2cm}
\includegraphics*[height=35mm]{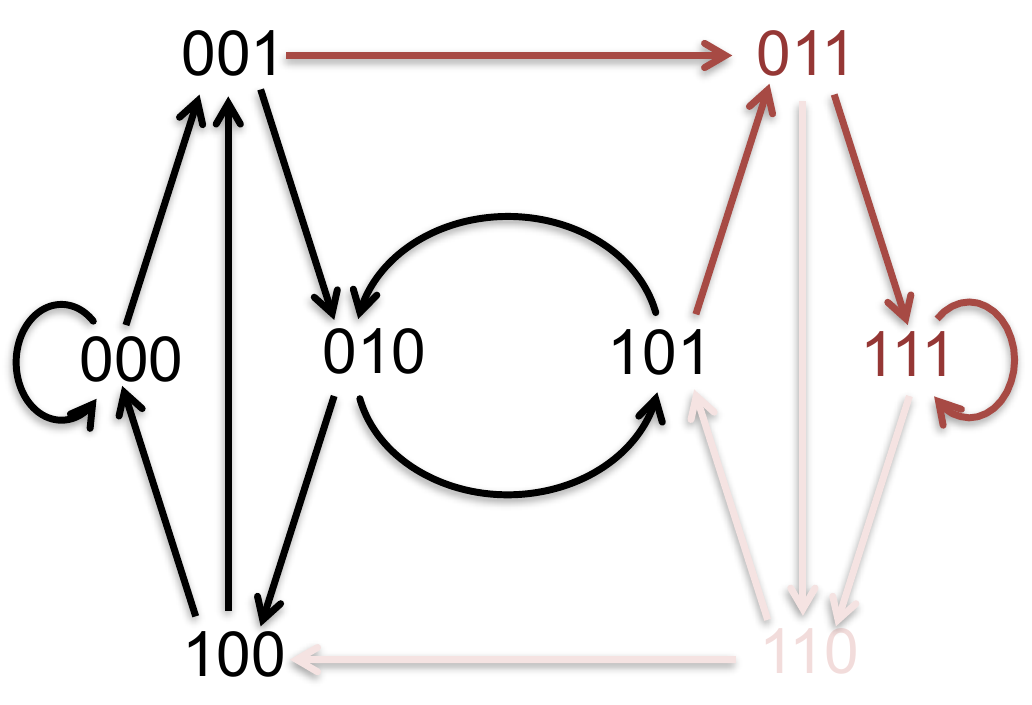}
\hspace{0.2cm}
\includegraphics*[height=35mm]{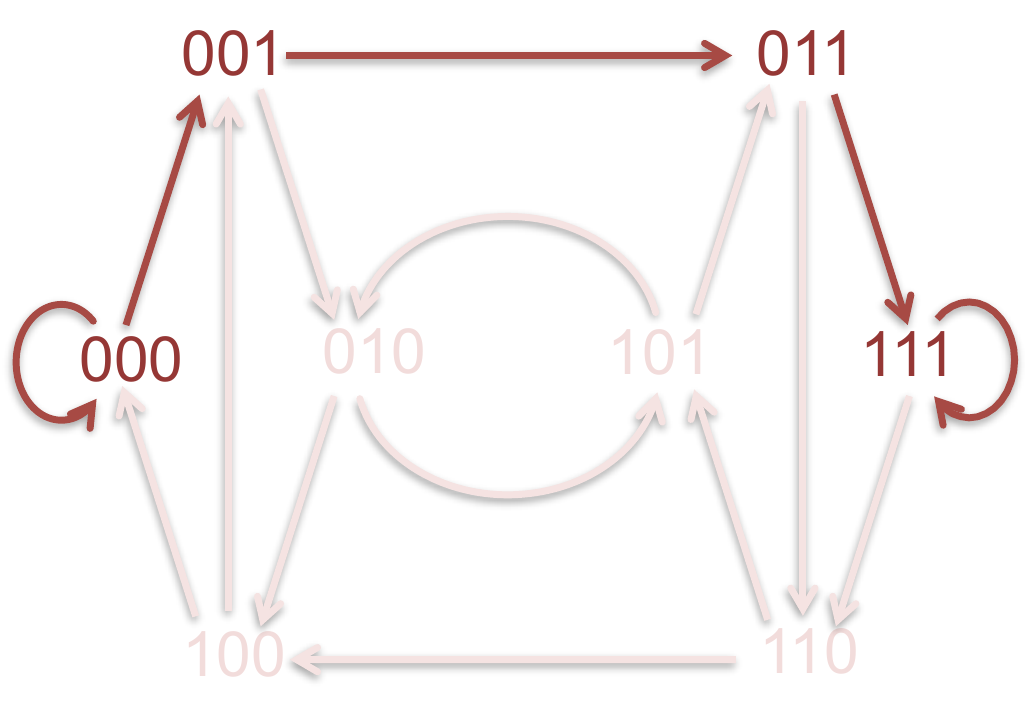}
\end{center}
\caption{Graphs associated with the (nested) 3-blocks subshifts that generate the (parameter dependent) fixed point sets of the cellular automaton ${\cal F}_2$. {\sl Left.} DeBruijn graph $B(2,3)$ (see Wikipedia or \cite{dB46}) associated with the full shift. {\sl Center.} All 3-blocks but $110$ are allowed. {\sl Right.} The four blocks $110$, $010$, $101$ and $100$ are forbidden. Light red color corresponds to blocks/transitions that are forbidden. Dark red color corresponds to transient blocks/transitions that do not contribute to the subshift entropy.}
\label{DIGRAPHRANK2}
\end{figure}

In order to determine the existence of fixed points ($v=0$), one needs to order the values of $X_2$ for $\theta^{1}=\theta^0$. Using the symmetry $X_2(1-\cdot)=1-X_2(\cdot)$, it suffices to consider those $\theta^{1}=\theta^0$ with $\theta_0^1=\theta_0^0=0$. Using for simplicity, only the top row to identify the 3-blocks in this case, one can show that the corresponding values comply with the lexicographic order, {\sl ie.}\ we have 
\[
0=X_2\left(000\right)<X_2\left(100\right)<X_2\left(010\right)<X_2\left(110\right).
\] 
Now, $\theta\in\{0,1\}^\Z$ is a fixed point of ${\cal F}_2$ iff the following constraint holds on its 3-blocks $\{\theta_{s-2}\theta_{s-1}\theta_s\}_{s\in\Z}$ 
\[
H\left(X_2\left(\theta_{s-2}\theta_{s-1}\theta_s\right)-T\right)=\theta_s,\ \forall s\in\Z.
\]
As mentioned above, this constraint is regarded as defining a subshift over 3-blocks, which depends on parameters, see Fig.\ \ref{DIGRAPHRANK2} for the graphs obtained from the analysis to follow. 

By considering the relative position of $T$ with respect to the ordered values of $X_2$ above, the following claims result about the existence of fixed point subshifts. Needless to say that similar conclusions immediately follow for the TW with $v=1$, from the symmetry $\epsilon\leftrightarrow 1-\epsilon$. 
\begin{itemize}
\item The full shift is admissible\footnote{This means that every element in this set if a fixed point.} iff $X_2\left(110\right)<T\leq 1-X_2\left(110\right)$.
\item The subshift for which all 3-blocks but $110$ are allowed is admissible iff\footnote{To see this, note that we have $H(X_2(\theta_{2}\theta_{1}0,\theta_{1}0)-T)=0$ except when $\theta_{2}=\theta_{1}=1$ and $H(X_2(\theta_{2}\theta_{1}1,\theta_{1}1)-T)=1$ for every $\theta_{2},\theta_{1}\in\{0,1\}$.}  
\[
X_2\left(010\right)<T\leq \min\left\{X_2\left(110\right),X_2\left(001\right)\right\}.
\]
By symmetry $T\leftrightarrow 1-T$, the subshift for which all 3-blocks but $011$ are allowed is admissible iff $1-T$ satisfies the same condition.
\item There are no other possibly admissible subshifts of positive entropy. In particular, the only non-homogeneous fixed points which exist for 
$0<T\leq \min\left\{X_2\left(010\right),X_2\left(001\right)\right\}$, 
write $\{H(s)\}\ \text{mod}\ \sigma$.\footnote{Indeed, we then have $H\left(\left(\begin{matrix}\theta10\\ \hfill 10\end{matrix}\right)-T\right)=1$ for $\theta\in\{0,1\}$. Alternatively, one first observes that the new constraint $T\leq X_2\left(010\right)$ immediately implies that $010$ is forbidden. The transition graph on Fig.\ \ref{DIGRAPHRANK2} then implies that the block $101$, and then also $100$, cannot be accessed either. The only non-trivial admissible path is the one that joins $000$ to $111$. Notice also that these fixed points  exist in the larger interval $0<T\leq \min\left\{X_2\left(110\right),X_2\left(001\right)\right\}$.} By symmetry $T\leftrightarrow 1-T$, the only non-trivial fixed points for $\max\left\{X_2\left(001\right),X_2\left(110\right)\right\}<T\leq 1$ are $\{1-H(s)\}\ \text{mod}\ \sigma$.
\end{itemize}
As before, the intervals in either of the first two items are non-empty iff $\epsilon$ is not too large (and smaller than some threshold smaller than $\tfrac12$ that depends on $a$). Moreover, they are adjacent when all non-empty. The intervals in the third claim are never empty, see Fig.\ \ref{APPROXRANK2}.

As entropy is concerned, it is obviously equal to $\log 2$ in the first case. To compute it in the second case (and also for any other subshift in the sequel), we consider that this quantity is not affected by passing to non-wandering sets, so that the transitions painted in red in the central picture of Fig.\ \ref{DIGRAPHRANK2} can be ignored. The remaining subshift turns out to be identical to the {\sl golden shift}, Fig.\ \ref{SYMBOLICGRAPHS} (a), whose entropy can be easily computed as the logarithm of the largest eigenvalue $\frac{1+\sqrt{5}}2\simeq 1.618$ of the matrix $\left(\begin{matrix}1&1\\1&0\end{matrix}\right)$.

\subsubsection{Analysis of TW with velocity $v=\tfrac12$}
\begin{figure}[ht]
\begin{center}
\includegraphics*[height=35mm]{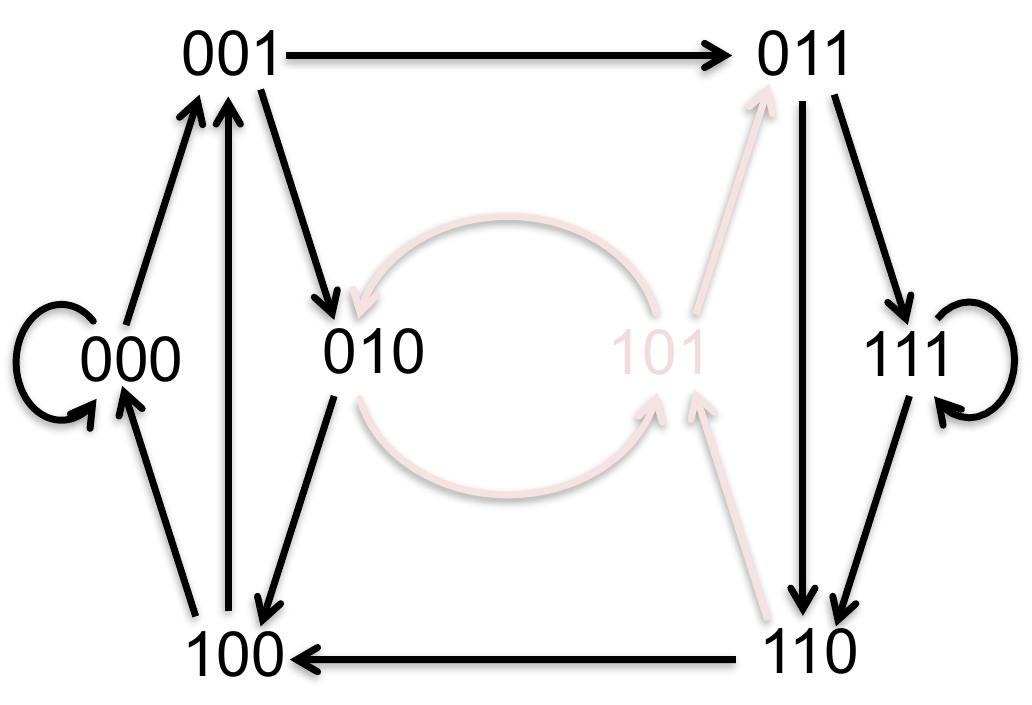}
\hspace{1cm}
\includegraphics*[height=35mm]{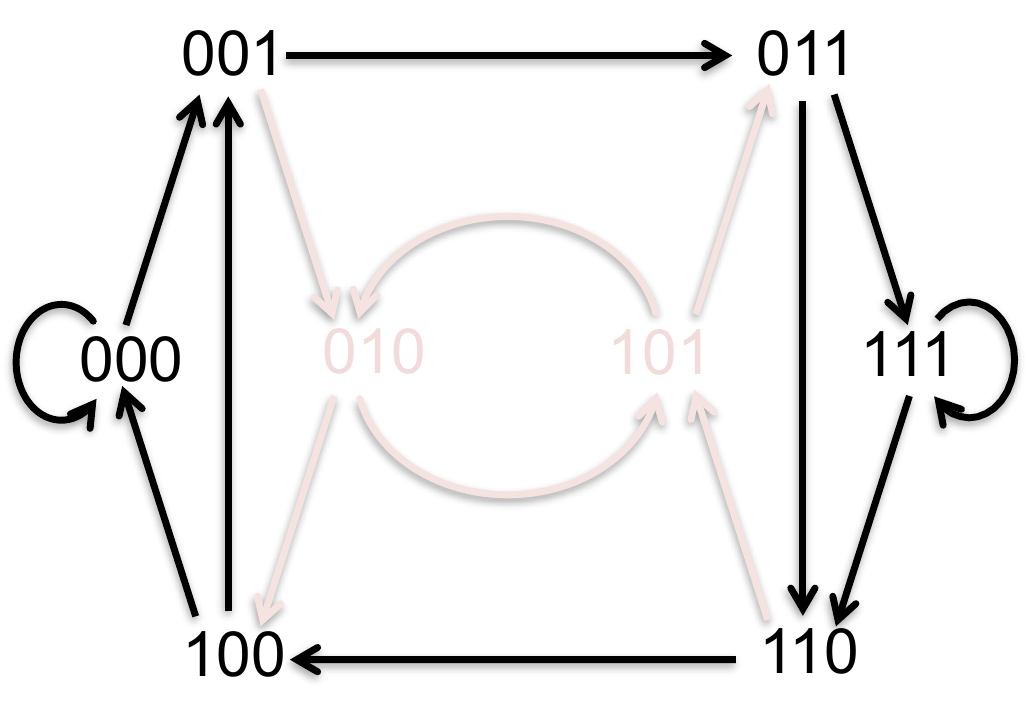}
\end{center}
\caption{Graphs associated with subshifts that generate TW of velocity $\tfrac12$ of the CA ${\cal F}_2$. {\sl Left.} The block $101$ is the only forbidden one. {\sl Right.} Both $101$ and $010$ are forbidden. Same color codes as in Fig.\ \ref{DIGRAPHRANK2}.}
\label{GRAPHSRK2V1/2}
\end{figure}

A configuration $\theta\in\{0,1\}^\Z$ of a TW with velocity $v=\tfrac12$ is defined by $\theta={\cal F}_2(\sigma^{-1}(\theta),\theta)$ (or equivalently by $\sigma(\theta)={\cal F}_2(\theta,\sigma(\theta))$). In this case, the analysis of admissible 3-blocks then amounts to consider both quantities $X_2\left( \begin{matrix}\theta_2\theta_1\theta_0\\ \hfill\theta_1\theta_0\end{matrix}\right)$ and $X_2\left( \begin{matrix}\theta_2\theta_1\theta_0\\ \hfill \theta_2\theta_1\end{matrix}\right)$ and to obtain conditions so that its output is equal to $\theta_1$ in both cases. The results are given below and the corresponding transition graphs are presented in Fig.\ \ref{GRAPHSRK2V1/2}. 
\begin{itemize}
\item The subshift for which all 3-blocks but $101$ are allowed (Fig.\ \ref{GRAPHSRK2V1/2} left) is admissible iff\footnote{Indeed, we then have $H\left( X_2\left( \begin{matrix}\theta_2\theta_1\theta_0\\ \hfill \theta_1\theta_0\end{matrix}\right)-T\right)=H\left(X_2\left( \begin{matrix}\theta_2\theta_1\theta_0\\ \hfill \theta_2\theta_1\end{matrix}\right)-T\right)=\theta_1$ for all $\theta_2\theta_1\theta_0\neq 101$. Conversely, that the block $101$ is not admissible is obvious from the condition on $T$. (NB: When the space-time symbol block corresponds to a fixed point, we use the simplified notation.)} 
\[
\max\left\{X_2\left(001\right),X_2\left( \begin{matrix}100\\ \hfill 10\end{matrix}\right)\right\}<T\leq \min\left\{X_2\left(010\right),X_2\left( \begin{matrix}010\\ \hfill 01\end{matrix}\right)\right\}.
\]
By the symmetry $T\leftrightarrow 1-T$, the subshit in which no 3-block is equal to $010$ is admissible iff $1-T$ satisfies the same condition (which amounts to reflect the interval wrt to $\tfrac12$).
\item The subshift for which all 3-blocks but $101$ and $010$ are allowed (Fig.\ \ref{GRAPHSRK2V1/2} right) is admissible iff 
\[
\max\left\{X_2\left(001\right),X_2\left( \begin{matrix}100\\ \hfill 10\end{matrix}\right)\right\}<T\leq \min\left\{X_2\left(110\right),X_2\left( \begin{matrix}011\\ \hfill 01\end{matrix}\right)\right\}.
\]
\item There are no other possibly admissible subshifts of positive entropy. In particular, no non-homogeneous TW with $v=\tfrac12$ exists when the previous condition fails.
\end{itemize}
As before, the intervals for $T$ depend on $\epsilon$, but this time they are non-empty only when $\epsilon$ is close enough to $\tfrac12$ (again depending on $a$), and complement existence domains of chaos of fixed points and TW with $v=1$, see Fig.\ \ref{APPROXRANK2}. Moreover, the interval in the second item contains the two ones in the first item.\footnote{Indeed, we have $\min\left\{X_2\left( \begin{matrix}010\\ \hfill 10\end{matrix}\right),X_2\left( \begin{matrix}010\\ \hfill 01\end{matrix}\right)\right\}<\min\left\{X_2\left( \begin{matrix}011\\ \hfill 01\end{matrix}\right),X_2\left( \begin{matrix}110\\ \hfill 10\end{matrix}\right)\right\}=1-\max\left\{X_2\left( \begin{matrix}100\\ \hfill 10\end{matrix}\right),X_2\left( \begin{matrix}001\\ \hfill 01\end{matrix}\right)\right\}$.} 
These two intervals never intersect,\footnote{because $\min\left\{X_2\left( \begin{matrix}010\\ \hfill 10\end{matrix}\right),X_2\left( \begin{matrix}010\\ \hfill 01\end{matrix}\right)\right\}\leq\tfrac12$} which means that it is impossible to have full chaos of TW with $v=\tfrac12$. 

\begin{figure}[ht]
\begin{center}
\includegraphics*[width=150mm]{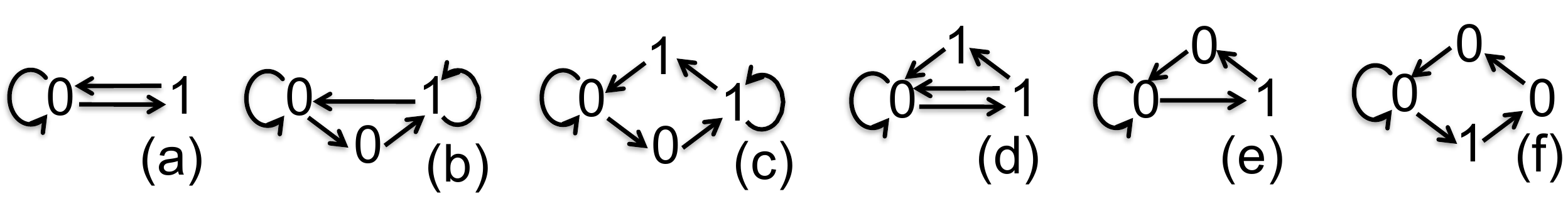}
\end{center}
\caption{Transitions graphs of some simple subshifts, which capture the entropy of TW profile subshifts in the CA ${\cal F}_R$, see text more details.}
\label{SYMBOLICGRAPHS}
\end{figure}

As entropy is concerned, using a similar non-wandering argument as above, we obtain that the entropy of the subshift(s) in the first item is obtained from the largest eigenvalue $\lambda\sim 1.755$ of the transition matrix $\left(\begin{matrix}1&1&0\\0&1&1\\1&0&0\end{matrix}\right)$ associated with the graph in Fig.\ \ref{SYMBOLICGRAPHS} (b). The entropy of the subshift in the second item is equal to that of the one depicted in Fig.\ \ref{SYMBOLICGRAPHS} (c), which turns out to be equal to that of the golden shift $\log\frac{1+\sqrt{5}}2\ (<\log 1.755)$.

\subsection{Rank 3 approximation}
\begin{figure}[ht]
\begin{center}
\includegraphics*[width=50mm]{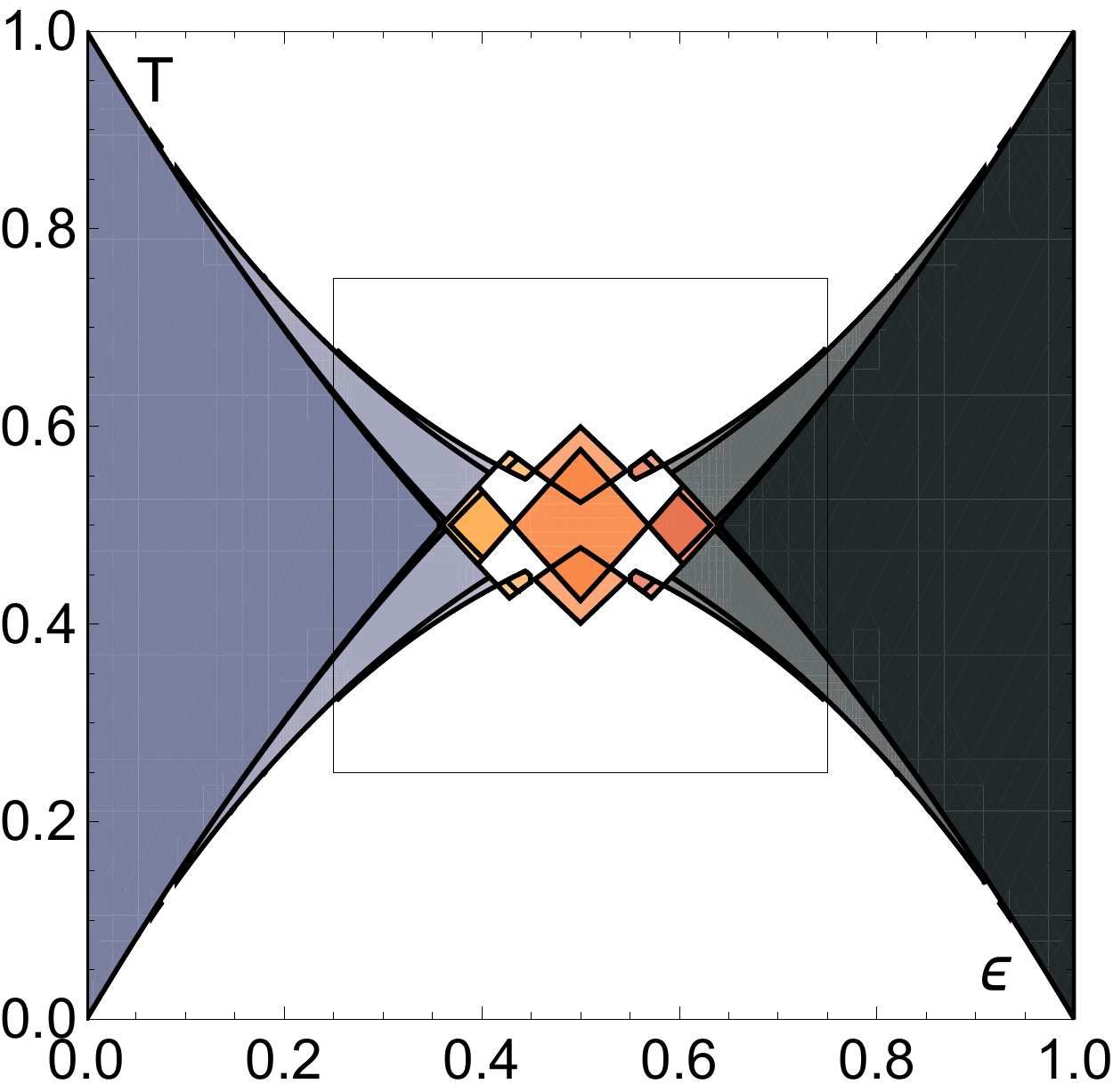}
\hspace{1.5cm}
\includegraphics*[width=50mm]{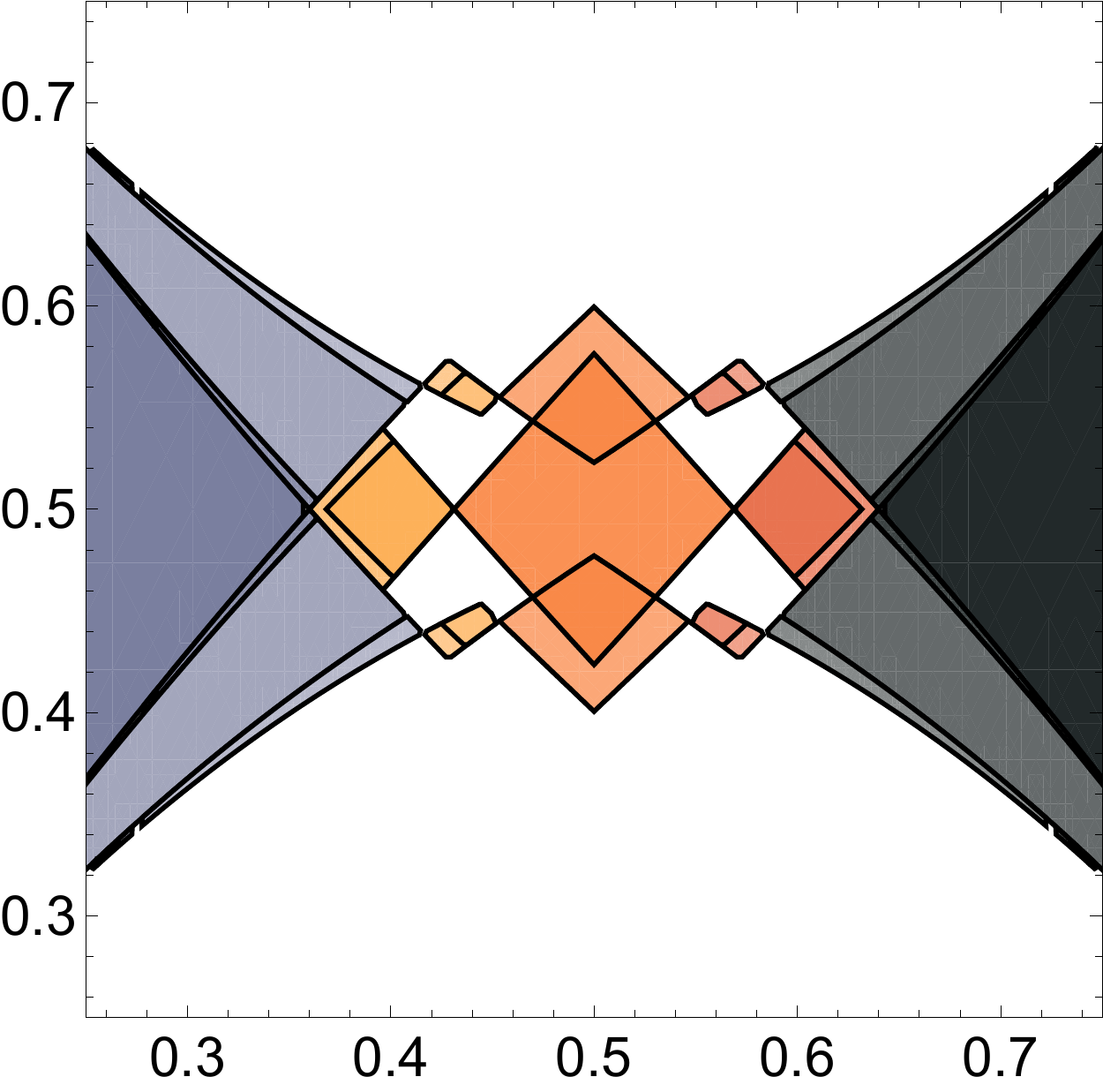}
\end{center}
\caption{Entropy-velocity diagram of TW of the cellular automaton ${\cal F}_3$ (and zoom into the central region). As before, chaos of fixed points is full in the central blue region  and the successive lower blue domains correspond to successive subshifts in Table 1/Fig.\ \ref{DIGRAPHRANK3}. The central orange domains is where we have chaos of TW with $v=\frac12$ (Table 2 and Fig.\ \ref{GRAPHSRK3V1/2}). The intermediate yellow and red domains respectively correspond to $v=\tfrac13$ and $v=\tfrac23$, see Table 3 and Fig.\ \ref{GRAPHSRK3V1/2} for a description of these domains, together with the corresponding subshifts.}
\label{APPROXRANK3}
\end{figure}
In order to obtain further refinement of the entropy-velocity diagram, the next logical step is to consider the rank 3 approximation ${\cal F}_3$. The analysis is similar to as before and we only provide the results here, which are illustrated by the entropy-velocity diagram in Fig.\ \ref{APPROXRANK3}. To that goal, we shall need to investigate the values of the function 
\[
X_3\left( \begin{matrix}\theta_3^2\theta_2^2\theta_1^2\theta_0^2\\
\hfill\theta_{2}^{1}\theta_{1}^{1}\theta_0^{1}\\
\hfill\theta_{1}^{0}\theta_0^{0}\end{matrix}\right)=\frac{1-a}{1-a^3}\sum_{k=0}^2 a^k\sum_{n=0}^{k+1}\ell_{n,k+1}\theta_{n}^{k},
\]

\begin{figure}[ht]
\begin{center}
\includegraphics*[height=28mm]{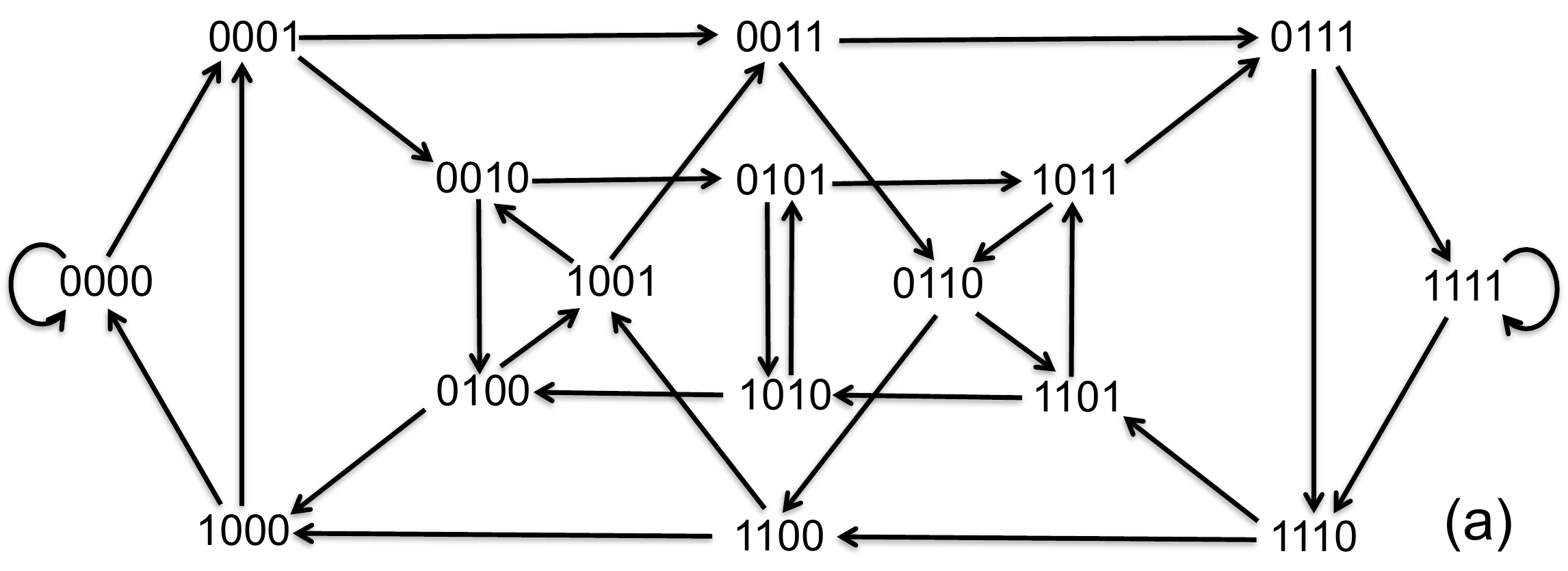}
\hspace{0.5cm}
\includegraphics*[height=28mm]{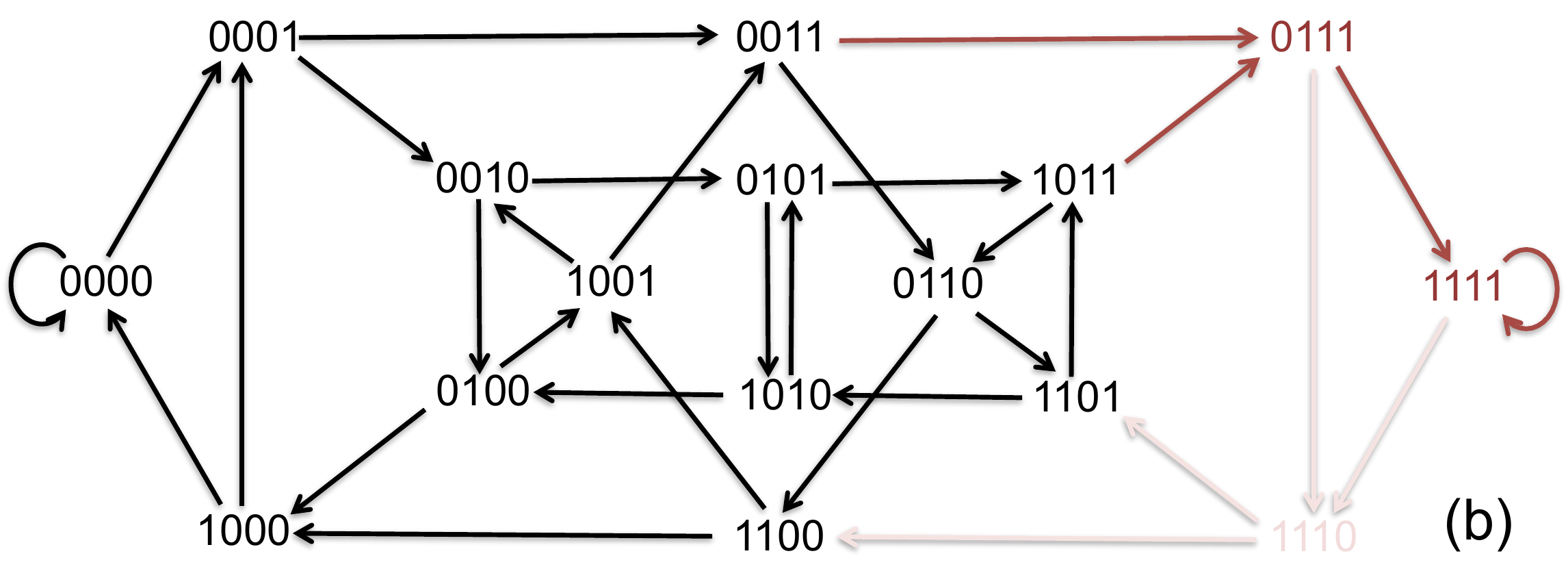}
\hspace{0.5cm}
\includegraphics*[height=28mm]{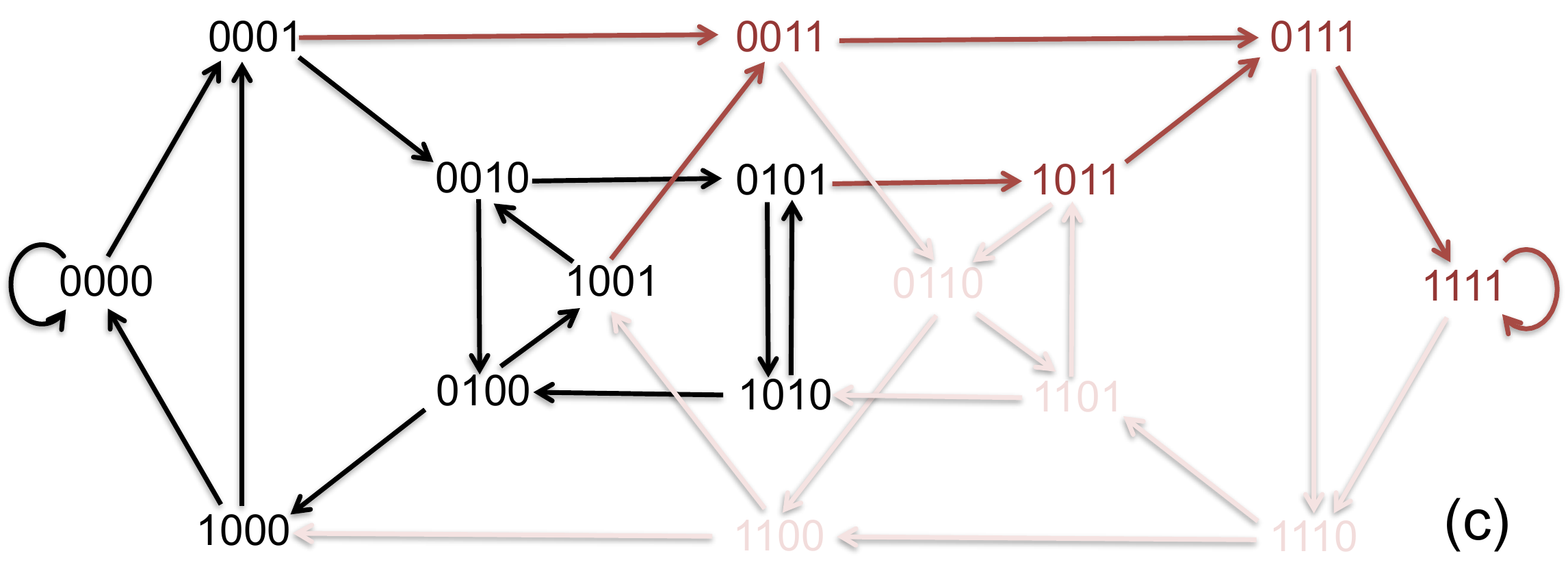}
\hspace{0.5cm}
\includegraphics*[height=28mm]{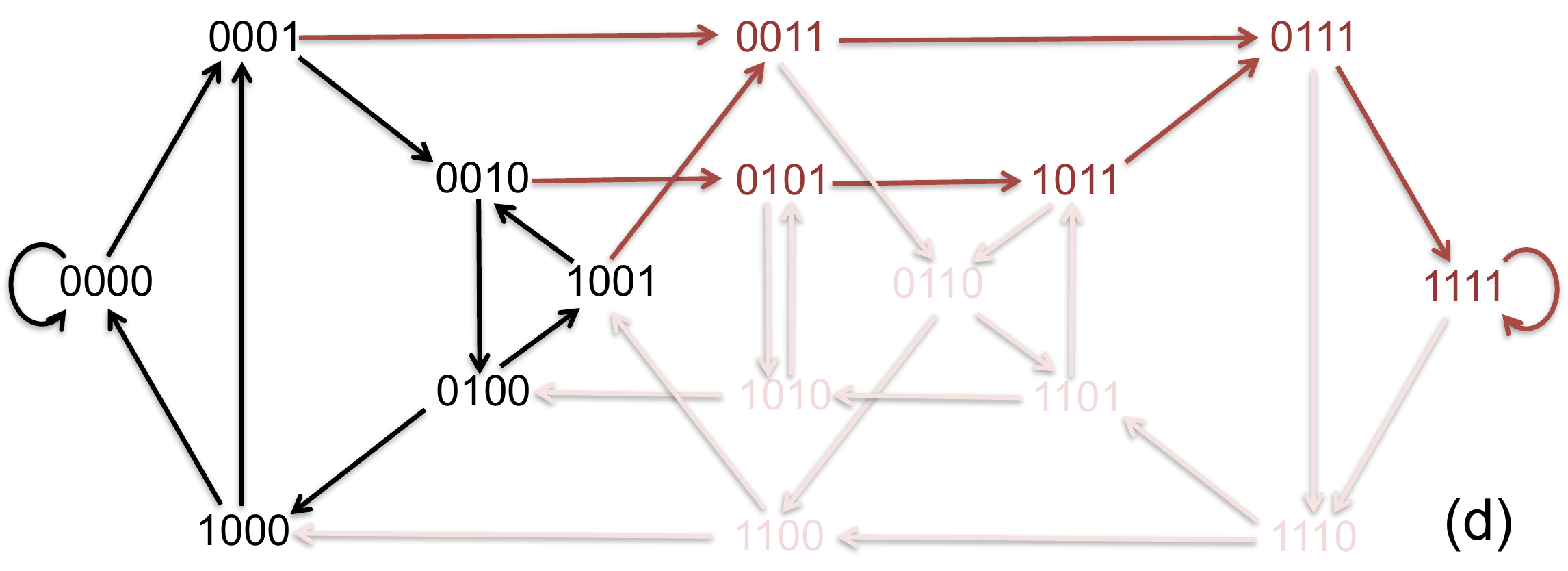}
\hspace{0.5cm}
\includegraphics*[height=28mm]{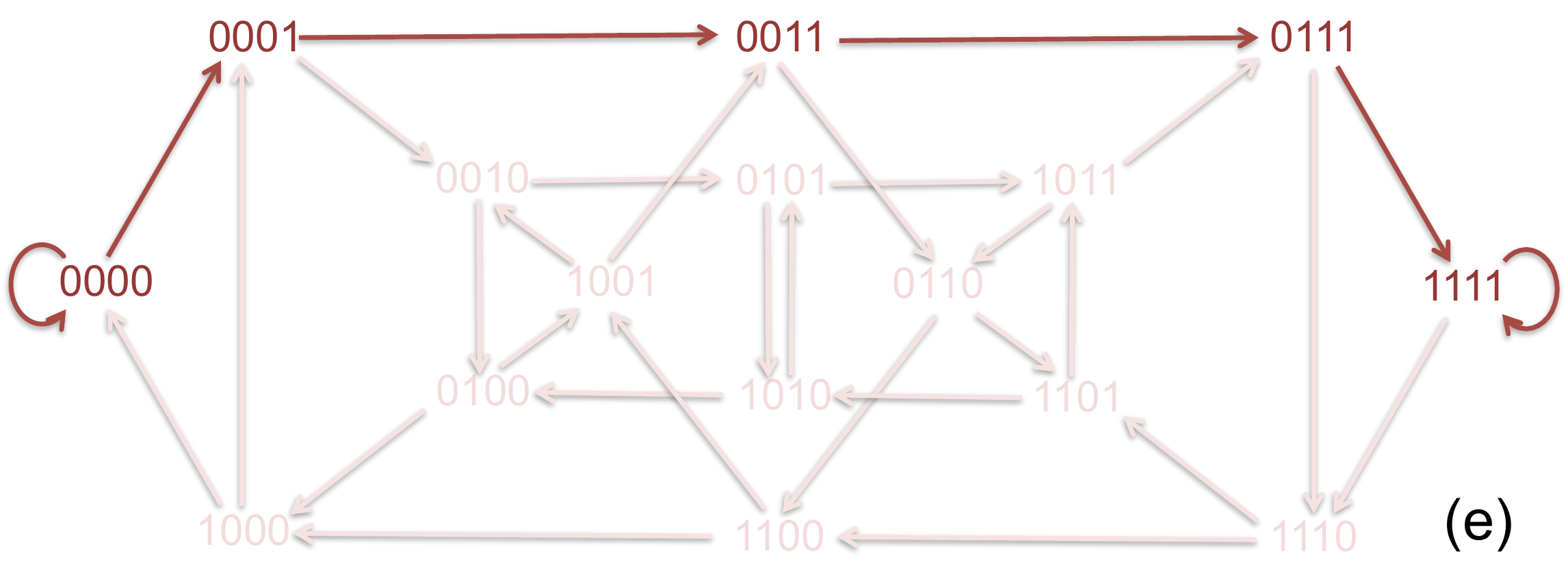}
\end{center}
\caption{Graphs associated with the (nested) 4-blocks subshifts in Table 1. Details as in Fig.\ \ref{DIGRAPHRANK2}.}
\label{DIGRAPHRANK3}
\end{figure}

\noindent
{\bf Fixed points.} 
As before, one can check that the values of this function, for configurations $\theta^2=\theta^1=\theta^0$ that are associated with fixed points, are ordered according the lexicographic order. We have for $\epsilon\in (0,\frac12)$, and using similar notation as for rank-2 fixed points,
\[
X_3(1000)<X_3(0100)<X_3(1100)<X_3(0010)<X_3(1010)<X_3(0110)<X_3(1110).
\]
The admissible 4-blocks subshifts depend on the relative location of the threshold $T$ with respect to this ordering. The possibly admissible subshifts are those listed in Table 1 - there are no other possibly admissible subshifts of positive entropy, excepted, evidently those obtained by the exchanging 0 and 1 - and illustrated in Fig.\ \ref{DIGRAPHRANK3}. 
\begin{table*}[!ht]
\begin{center}
\begin{tabular}{|c|c|c|c|}
\hline
Forbidden 4-blocks&Graphs&Entropy&Existence condition\\
\hline
None&Fig.\ \ref{DIGRAPHRANK3} (a)&$\log 2$&$X_3(1110)<T\leq X_3(0001)=1-X_3(1110)$\\
\hline
$1110$&Fig.\ \ref{DIGRAPHRANK3} (b)/\ref{SYMBOLICGRAPHS} (d)&$\log 1.839$&$X_3(0110)<T\leq \min\left\{X_3(1110),X_3(0001)\right\}$\\
\hline
$+$ $0110$, $1100$, $1101$&Fig.\ \ref{DIGRAPHRANK3} (c)/\ref{SYMBOLICGRAPHS} (a)&$\log\frac{1+\sqrt{5}}2$&$X_3(1010)<T\leq \min\left\{X_3(0110),X_3(0001)\right\}$\\
\hline
$+$ $1010$&Fig.\ \ref{DIGRAPHRANK3} (d)/\ref{SYMBOLICGRAPHS} (e)&$\log 1.466$&$X_3(0010)<T\leq \min\left\{X_3(1010),X_3(0001)\right\}$\\
\hline
all non-homogenous&Fig.\ \ref{DIGRAPHRANK3} (e)&0&$0<T\leq \min\left\{X_3(0010),X_3(0001)\right\}$\\
\hline
\end{tabular}
\end{center}
\caption{Summary of possibly admissible (nested) subshifts of fixed points in the CA ${\cal F}_3$, together with references to the graphs in Fig.\ \ref{DIGRAPHRANK3}, to their simplification in Fig.\ \ref{SYMBOLICGRAPHS} (where it applies), to entropy estimates and to the intervals of the parameter $T$ where they are admissible. (NB: The constraint in the first column means that words that are {\sl not} forbidden are all admissible. The symbol $+$ means {\sl in addition to forbidden blocks in the previous row}.)}
\end{table*}

As for the rank-2 fixed points, the intervals in $T$ are only non-empty when $\epsilon$ is not too large. When non-empty, these intervals are adjacent to each other. The subshift are nested, with decreasing entropy, when $T$ moves away from $\frac12$.

\begin{figure}[ht]
\begin{center}
\includegraphics*[height=28mm]{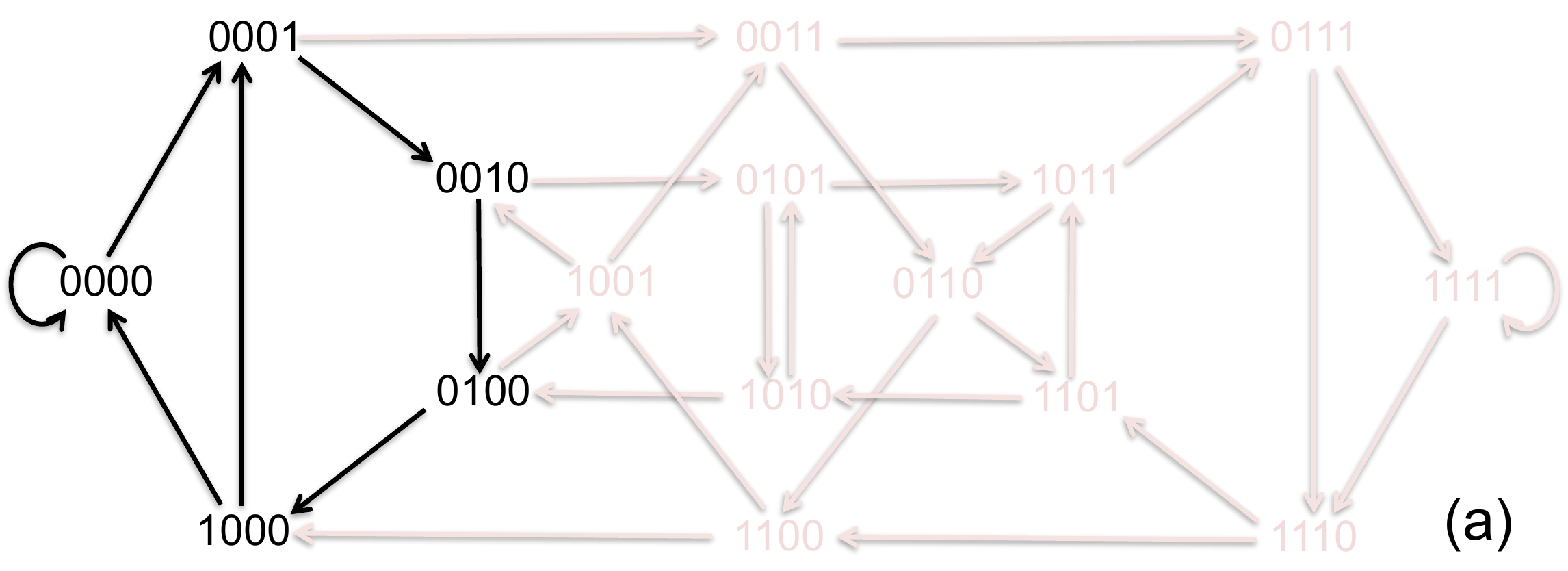}
\hspace{0.5cm}
\includegraphics*[height=28mm]{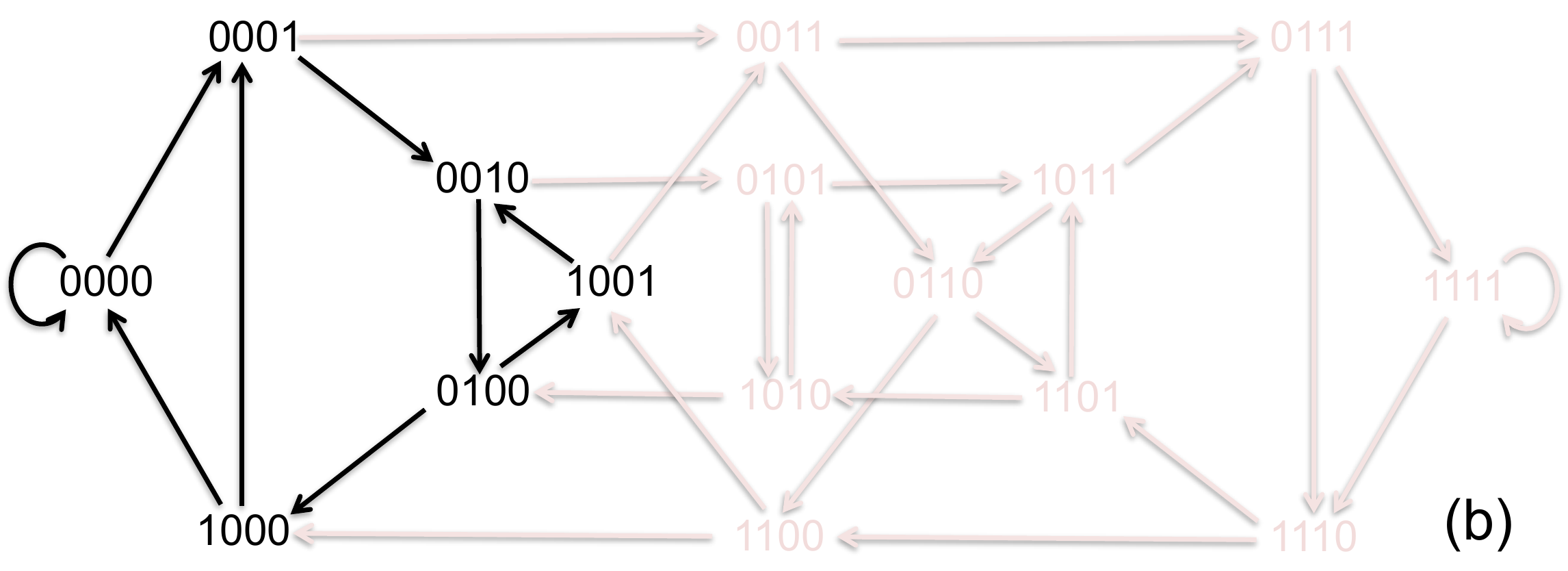}
\hspace{0.5cm}
\includegraphics*[height=28mm]{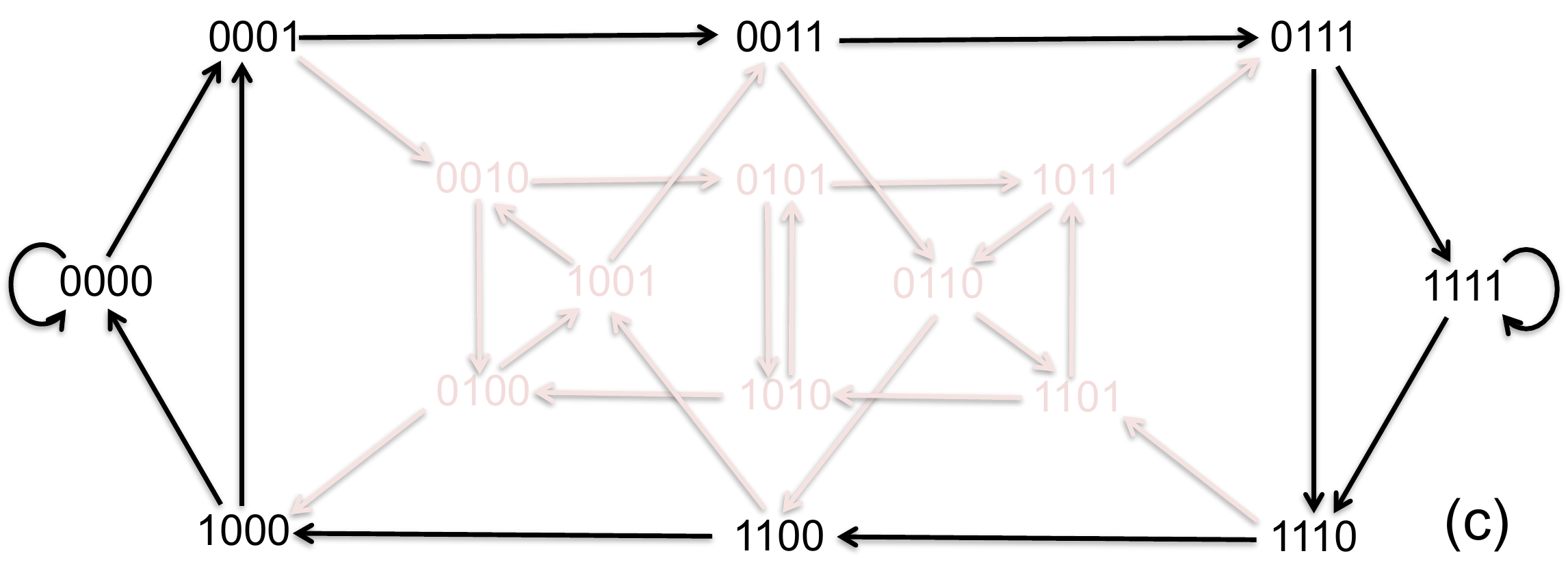}
\hspace{0.5cm}
\includegraphics*[height=28mm]{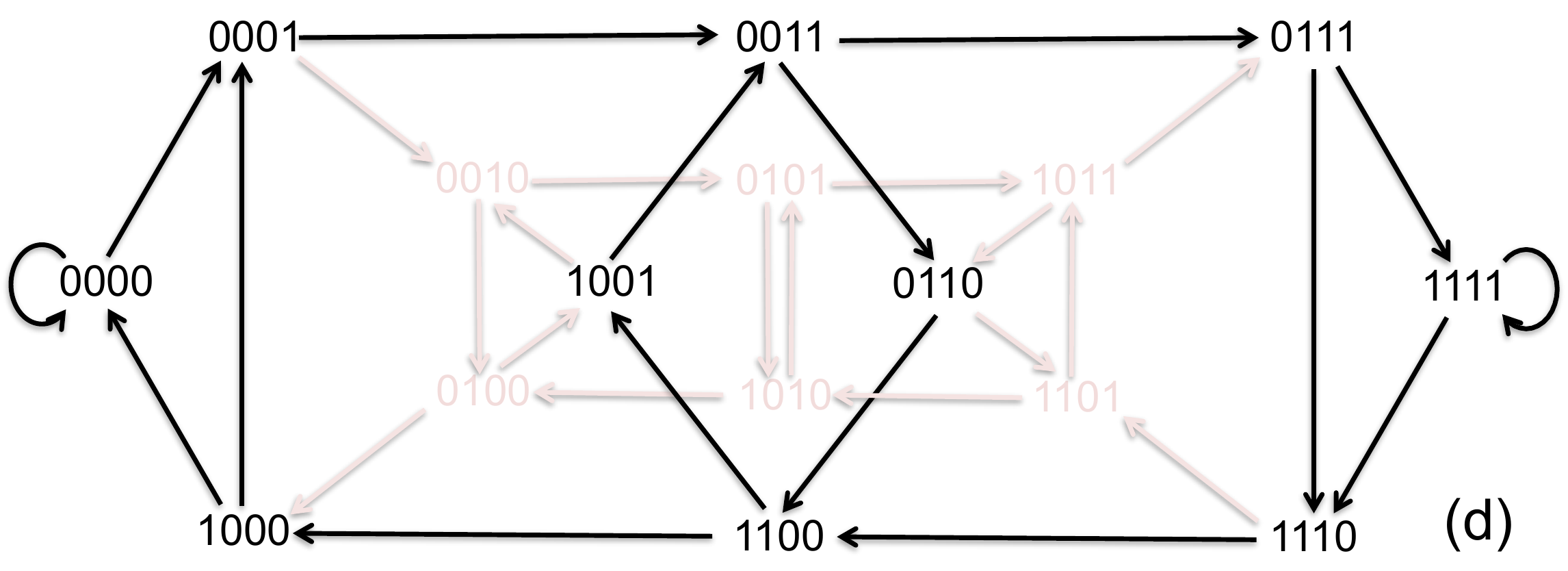}
\hspace{0.5cm}
\includegraphics*[height=28mm]{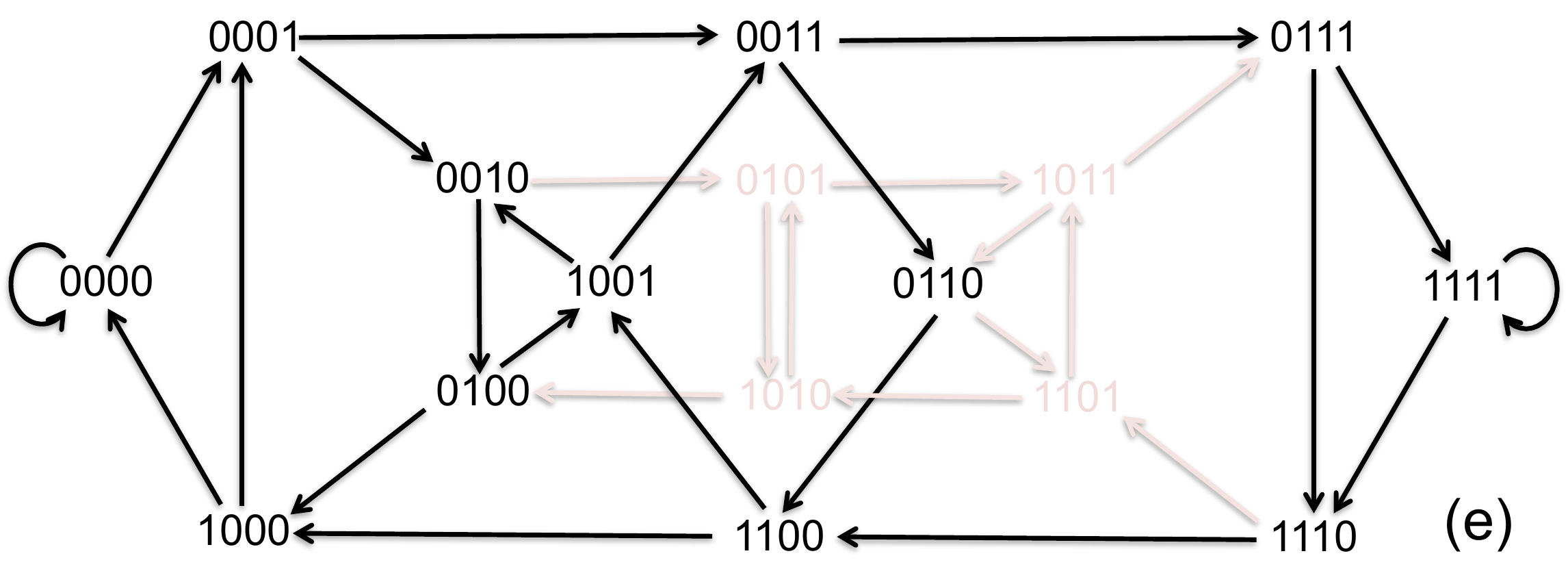}
\end{center}
\caption{Graphs associated with the subshifts in Table 2 and 3.}
\label{GRAPHSRK3V1/2}
\end{figure}

\noindent
{\bf TW with velocity $v=\tfrac12$.} 
For $v=\tfrac12$ TW, one needs to check that we simultaneously have 
\[
X_3\left( \begin{matrix}\theta_3\theta_2\theta_1\theta_0\\
\hfill\theta_{2}\theta_{1}\theta_0\\
\hfill\theta_{2}\theta_1\end{matrix}\right)=\theta_1\quad \text{and}\quad
X_3\left( \begin{matrix}\theta_3\theta_2\theta_1\theta_0\\
\hfill\theta_{3}\theta_{2}\theta_1\\
\hfill\theta_{2}\theta_1\end{matrix}\right)=\theta_2,
\]
for every $4$-block in the presumed subshifts. Candidate subshifts can be obtained from the 4-block graph associated with the full shift on 2 symbols, de Bruijn graph $B(2,4)$, Fig.\ \ref{DIGRAPHRANK3} (a). The possibly admissible subshifts are listed in Table 2  - there are no other possibly admissible subshifts of positive entropy, excepted, \footnote{In particular, the analysis shows that no subshift can be admissible without $1001$ and $0110$ being admissible. Similarly, the numerical computations of the existence conditions show that no admissible subshift can allow for $0101$ or $1010$.} evidently, those obtained by the exchanging 0 and 1 - and illustrated in Fig.\ \ref{GRAPHSRK3V1/2}.
\begin{table*}[!ht]
\begin{center}
\begin{tabular}{|c|c|c|}
\hline
Graphs&Entropy&Existence condition\\
\hline
Fig.\ \ref{GRAPHSRK3V1/2} (b)/\ref{SYMBOLICGRAPHS} (e)&$\log 1.446$&$\max\left\{X_3\left( \begin{matrix}0100\\
\hfill100\\
\hfill10\end{matrix}\right),X_3\left( \begin{matrix}0010\\
\hfill001\\
\hfill10\end{matrix}\right)\right\}<T\leq \min\left\{X_3\left( \begin{matrix}0010\\
\hfill010\\
\hfill01\end{matrix}\right),X_3\left( \begin{matrix}0100\\
\hfill010\\
\hfill10\end{matrix}\right)\right\}$\\
\hline
Fig.\ \ref{GRAPHSRK3V1/2} (d)/\ref{SYMBOLICGRAPHS} (c)&$\log \frac{1+\sqrt{5}}2$&$\max\left\{X_3\left( \begin{matrix}0011\\
\hfill001\\
\hfill01\end{matrix}\right),X_3\left( \begin{matrix}1100\\
\hfill100\\
\hfill10\end{matrix}\right)\right\}<T \leq \min\left\{X_3\left( \begin{matrix}1100\\
\hfill110\\
\hfill10\end{matrix}\right),X_3\left( \begin{matrix}0011\\
\hfill011\\
\hfill01\end{matrix}\right)\right\}$\\
\hline
Fig.\ \ref{GRAPHSRK3V1/2} (e)&$\log 1.674$&intersection of previous conditions\\
\hline
\end{tabular}
\end{center}
\caption{Summary of possibly admissible subshifts of TW $v=\tfrac12$ in the CA ${\cal F}_3$, together with references to the graphs in Fig.\ \ref{GRAPHSRK3V1/2}, to their simplification in Fig.\ \ref{SYMBOLICGRAPHS} (where it applies), to entropy estimates and to existence domains, expressed in terms of intervals for the parameter $T$. In particular, the intervals in the middle row are symmetric wrt $\frac12$.}
\end{table*}

As for the rank-2 TW $v=\frac12$, the intervals in $T$ are non-empty when $\epsilon$ is close to $\frac12$. However, when both non-empty, the intervals in the first and second rows overlap and enhancement of the entropy results when either $T$ or $1-T$ lies in their intersection  (Notice also, that the interval in the first row never intersects its reflected image wrt $\tfrac12$).

\noindent
{\bf TW with velocity $v=\tfrac13$.} 
In addition to fixed points and $v=\tfrac12$ TW, the rank 3 approximation may also have $v=\tfrac13$ (and $v=\frac23$) TW. We focus on $v=\tfrac13$ since the other one can be deduced from the symmetry $\epsilon\leftrightarrow 1-\epsilon$. In this case, a 4-block $\theta_3\theta_2\theta_1\theta_0$ is admissible iff the following three conditions simultaneously hold
\[
X_3\left( \begin{matrix}\theta_3\theta_2\theta_1\theta_0\\
\hfill\theta_{2}\theta_{1}\theta_0\\
\hfill\theta_{1}\theta_0\end{matrix}\right)=X_3\left( \begin{matrix}\theta_3\theta_2\theta_1\theta_0\\
\hfill\theta_{2}\theta_{1}\theta_0\\
\hfill\theta_{2}\theta_1\end{matrix}\right)=X_3\left( \begin{matrix}\theta_3\theta_2\theta_1\theta_0\\
\hfill\theta_{3}\theta_{2}\theta_1\\
\hfill\theta_{2}\theta_1\end{matrix}\right)=\theta_1,
\]
As before, candidate subshifts can be obtained by pruning the de Bruijn graph $B(2,4)$. Up to the symmetry $0\leftrightarrow 1$, the resulting possibly admissible subshifts are listed in Table 3 (again, no other subshift of positive entropy can be admissible), where the existence conditions have been simplified by assuming $\epsilon\leq \tfrac12$. The corresponding existence domains do not intersect those of previous velocities, except for some boundaries (complementary domains). The corresponding intervals in $T$ are only non-empty when $\epsilon$ lies in a intermediate range, between existence domains of velocities 0 and $\tfrac12$. The subshifts are similar to those associated with $v=\tfrac12$ TW (NB: in particular, no admissible subshift can allow for $0101$ or $1010$); however, there are two significant differences
\begin{itemize}
\item The admissibility of the blocks $1001$ and $0110$ is not a consequence of that of other blocks; hence there are additional (nested) existence domains (see 2nd and 4th rows in Table 3).
\item The existence domains do not overlap as they did; so that there no enhancement effect on the entropy exists in this case. 
\end{itemize}
\begin{table*}[!ht]
\begin{center}
\begin{tabular}{|c|c|c|}
\hline
Graphs&Entropy&Existence condition\\
\hline
Fig.\ \ref{GRAPHSRK3V1/2} (a)/\ref{SYMBOLICGRAPHS} (f)&$\log 1.380$&$\max\left\{X_3\left( 0001\right),X_3\left( \begin{matrix}0100\\
\hfill010\\
\hfill10\end{matrix}\right)\right\}<T\leq \min\left\{X_3\left(0010\right),X_3\left( \begin{matrix}0010\\
\hfill001\\
\hfill01\end{matrix}\right)\right\}$\\
\hline
Fig.\ \ref{GRAPHSRK3V1/2} (b)/\ref{SYMBOLICGRAPHS} (e)&$\log 1.446$&$\max\left\{X_3\left( 1001\right),X_3\left( \begin{matrix}0100\\
\hfill010\\
\hfill10\end{matrix}\right)\right\}<T\leq \min\left\{X_3\left(0010\right),X_3\left( \begin{matrix}0010\\
\hfill001\\
\hfill01\end{matrix}\right)\right\}$\\
\hline
Fig.\ \ref{GRAPHSRK3V1/2} (c)&$\log 1.466$&$\max\left\{X_3\left( 0001\right),X_3\left( \begin{matrix}1100\\
\hfill110\\
\hfill10\end{matrix}\right)\right\}<T \leq \min\left\{X_3\left(1110\right),X_3\left( \begin{matrix}0011\\
\hfill001\\
\hfill01\end{matrix}\right)\right\}$\\
\hline
Fig.\ \ref{GRAPHSRK3V1/2} (d)/\ref{SYMBOLICGRAPHS} (c)&$\log \frac{1+\sqrt{5}}2$&$\max\left\{X_3\left( 1001\right),X_3\left( \begin{matrix}1100\\
\hfill110\\
\hfill10\end{matrix}\right)\right\}<T \leq \min\left\{X_3\left(0110\right),X_3\left( \begin{matrix}0011\\
\hfill001\\
\hfill01\end{matrix}\right)\right\}$\\
\hline
\end{tabular}
\end{center}
\caption{Summary of possibly admissible subshifts of TW $v=\tfrac13$ in the CA ${\cal F}_3$, together with references to the graphs in Fig.\ \ref{GRAPHSRK3V1/2}, their simplification in Fig.\ \ref{SYMBOLICGRAPHS} (where it applies), entropy estimates and existence domains. The intervals in the last two rows are symmetric wrt to $\frac12$. The interval in the 2nd (resp.\ 4th) row is contained in the one of the 1st (resp.\ 3rd) row. The entropy of the subshift in rows 2 and 3 are unexpectedly equal.}
\end{table*}

\section{Concluding remarks}
The analysis above has revealed that chaos of TW is selective in basic finite rank approximations ${\cal F}_R$ of the LDS $F_\epsilon$, with unique velocity depending on parameters. More than in $F_\epsilon$, in the ${\cal F}_R$, this uniqueness applies to all velocities and not only to $v\in (0,1)$. For ${\cal F}_3$, some boundaries curves in the $(\epsilon,T)$ square, of the domains associated with $v=\tfrac13$, coincide with some boundaries associated with $v=0$ (Fig.\ \ref{APPROXRANK3}) (and a similar coincidence holds for $v=\tfrac12$ and $v=0$ in Fig.\ \ref{APPROXRANK2}). This suggests that some overlap between the domains for $v=0$ and $v=\tfrac1{R}$ might exist for $R$ large enough.
 
Moreover, even though this analysis is too preliminary to anticipate full results for larger values of $R$, some systematic features have emerged. In particular, the fact that for any $R$, the following {\sl extreme} subshifts can be certainly be admissible, depending on the parameters:
\begin{itemize}
\item $B_R$: every block of consecutive 0's (or consecutive 1's) must be of length $R$ or longer; a generalisation of the graphs in Fig.\ \ref{GRAPHSRK2V1/2} right and Fig.\ \ref{GRAPHSRK3V1/2} (c).
\item $A_R$: every block of consecutive 0's must be of length $R$ or larger, and every 1 must be isolated (or the subshift resulting from exchanging 0's and 1's); a generalisation of the graphs in Fig.\ \ref{GRAPHSRK3V1/2} (a).
\end{itemize}
Sufficient existence conditions for these subshifts have been provided in \cite{FLU09}, in term of existence regions of fronts and solitary waves respectively. Namely, domains in the $(\epsilon,T)$ square have been given, so that the subshift $B_R$ (resp.\ $A_R$) of TW with $v=\frac{p}{q}$ given and $R$ sufficiently large, is admissible. However, to compute existence  conditions of $A_R$ and $B_R$ for any $R$, is not totally obvious, even though there are some similarities between the case $R=2$ and $R=3$. Those computations could be part of a continuation to this paper. 
\bigskip

\noindent
{\bf Acknowledgements.} I am grateful to Stanislav M. Mintchev for careful reading of the manuscript, comments and suggestions.


\end{document}